\documentclass[fleqn,usenatbib]{mnras}

\usepackage{newtxtext,newtxmath}

\usepackage[T1]{fontenc}

\DeclareRobustCommand{\VAN}[3]{#2}
\let\VANthebibliography\thebibliography
\def\thebibliography{\DeclareRobustCommand{\VAN}[3]{##3}\VANthebibliography}

\newcommand{\Sint}{{S_{\rm int}}}
\newcommand{\Speak}{{S_{\rm peak}}}
\newcommand{\boot}{Bo\"{o}tes }
\newcommand{\qtir}{q_{\rm TIR}}
\newcommand{\ltir}{L_{\rm TIR}}
\newcommand{\tdust}{T_{\rm dust}}

\usepackage{graphicx}	
\usepackage{amsmath}	
\usepackage{soul}

\title[\boot field at 400\,MHz]{Detailed study of the \boot field using 300-500\,MHz uGMRT observations: Source Properties \& radio--infrared correlations}

\author[Akriti Sinha et al.]{
Akriti Sinha,\thanks{E-mail: sinha.akriti44@gmail.com}
Abhirup Datta,
\\
$^{1}$Indian Institute of Technology Indore, India
}

\date{Accepted XXX. Received YYY; in original form ZZZ}

\pubyear{2015}

\begin{document}
\label{firstpage}
\pagerange{\pageref{firstpage}--\pageref{lastpage}}
\maketitle

\begin{abstract}

The dominant source of radio continuum emissions at low frequencies is synchrotron radiation, which originates from star-forming regions in disk galaxies and from powerful jets produced by active galactic nuclei (AGN). We studied the \boot field using the upgraded Giant Meterwave Radio Telescope (uGMRT) at 400\,MHz, achieving a central minimum off-source RMS noise of 35\,$\mu$Jy\,beam$^{-1}$ and a catalogue of 3782 sources in $\sim6$ sq. degrees of the sky. The resulting catalogue was compared to other radio frequency catalogues, and the corrected normalised differential source counts were derived. We use standard multi-wavelength techniques to classify the sources in star-forming galaxies (SFGs), radio-loud (RL) AGN, and radio-quiet (RQ) AGN that confirm a boost in the SFGs and RQ\,AGN AGN populations at lower flux levels. For the first time, we investigated the properties of the radio--IR relations at 400\,MHz in this field. The $L_{\rm 400\,MHz}$--$\ltir$ relations for SFGs were found to show a strong correlation with non-linear slope values of $1.10\pm0.01$, and variation of $\qtir$ with $z$ is given as, $q_{\rm TIR} = (2.19 \pm 0.07)\ (1+z)^{-0.15 \pm 0.08}$. This indicates that the non-linearity of the radio--IR relations can be attributed to the mild variation of $\qtir$ values with $z$. The derived relationships exhibit similar behaviour when applied to LOFAR at 150\,MHz and also at 1.4\,GHz. This emphasises the fact that other parameters like magnetic field evolution with $z$ or the number densities of cosmic ray electrons can play a vital role in the mild evolution of $q$ values.

\end{abstract}




\section{Introduction}

Unaffected by dust obscuration, radio emissions are important for diverse galaxy population studies and, thus, to probe the astrophysical processes at high redshifts. The synchrotron radiation is the emission mechanism responsible for the radio emissions from the relativistic jets powered by AGN and from the relativistic electrons accelerated by the supernova (SN) explosions in SFGs.
The emission of radio waves from star-forming galaxies may serve as a reliable tracer of the star formation rate \citep[SFR;][]{Yun_2001,Bell_2003,Pannella_2009}. As the SFR estimated at radio frequencies is unaffected by dust extinction, it is unbiased when compared to UV or visible wavelengths.

Since radio emissions can be used as a tracer of SFR, there lies a strong, near-linear correlation between radio and infrared (IR) luminosities yielding one of the tightest correlations observed in astrophysics, the \textit{radio-IR} relations \citep{Helou_1985,Condon_1992,Sargent_2010_2}. It is believed that the correlation between radio and IR emissions is due to the presence of massive ($\gtrsim10\,\rm M_\odot$) OB-type stars that emit ultraviolet (UV) photons. 
Subsequent to the absorption of UV photons, dust particles re-emit radiation in the IR waveband.
Meanwhile, when they reach the end of their short lives (a few million years), they explode as supernovae, providing synchrotron-emitting cosmic ray electrons (CREs).

With deep observations of the extra-galactic radio sky at low frequencies, one can probe the properties of the source population up to the early Universe \citep[see][]{Condon_1984,Becker1995,Gruppioni_1999}. The deep radio sky consists of distinct populations ranging from SFGs through faint AGN to RL\,AGN observed over various flux density regimes. The ongoing surveys, for instance, the VLASS using the Karl G. Jansky Very Large Array \citep[VLA;][]{Lacy_2020}, the MIGHTEE survey using the MeerKAT \citep{Jarvis_2016}, and the LoTSS from the Low-Frequency Array \citep[LOFAR;][]{Shimwell_2017,Tasse_2021} have opened up a new window for the extra-galactic deep radio-continuum observations at micro-Jansky levels ($\upmu$Jy). In fact, the advent of the surveys using the Square Kilometre Array \citep[SKA;][]{Jarvis_2015SK} will observe and detect millions of radio sources at unprecedented levels, as well as the next-generation VLA \citep[ngVLA;][]{Francesco_2019}.

Moreover, the radio sky is dominated by RL\,AGN at higher flux densities, while at fainter flux densities ($<$1\,mJy at 1.4\,GHz), the source population is dominated by SFGs \citep[see for eg.][]{Smolcic_2008,Padovani_2016,Arnab2019}. In the studies conducted by \cite{Padovani_2009} and \cite{Bonzini_2013}, a previously unknown group of sources called the RQ\,AGN was discovered at lower flux densities. Despite having low flux values, these sources display indicators of AGN activity in at least one of the electromagnetic spectrum bands.
\cite{Miller_1993} proposed that RQ\,AGN are scaled versions of RL\,AGN, whereas, \cite{Sopp_1991} determined that the radio emissions detected in these systems are the result of star formation processes occurring within the host galaxy.
To distinguish between the radio emission from AGN or SFGs based on radio flux density alone is not straightforward, which thus emphasises the need for a multi-wavelength study. According to \cite{Padovani_2011}, for a comprehensive classification of sources, a combination of X-ray, IR, and radio data is essential. \cite{Bonzini_2013} further offered an updated classification criterion in the Extended Chandra Deep Field-South (ECDFS).

Similarly, many radio deep fields have been exploited to study the source properties using the wealth of ancillary multi-wavelength data available, such as the Cosmic Evolution Survey (COSMOS) field \citep{Smolcic_2017}, the Lockman Hole (LH) field \citep{Prandoni_2018,Bonato2021}, the European Large-Area \textit{ISO} Survey-North 1 \citep[ELAIS-N1;][]{Ocran_2017,Ocran_2019,Akriti_2022}, along with others. However, there are still a lot of unanswered questions on this subject, and there is no universal agreement overall. Analyses of newer samples are required to determine the impact of various classification criteria and enhance the statistics further.

Among others, the \boot field is recognised as one of the largest and most extensively investigated extra-galactic fields in the sky. The NOAO Deep Wide Field Survey \citep[NDWFS;][]{Jannuzi_1999} originally observed the \boot field as part of its survey, which covered approximately 9\,deg$^2$ in the optical ($B_{\rm W}$, $R$, and $I$) and near-infrared ($K$) bands. Besides, the field also contains plenty of other multi-wavelength data that includes mid-infrared \citep{Eisenhardt_2004}, X-ray \citep{Murray_2005,Kenter2005}, and UV \citep[GALEX;][]{Martin_2003} wavelengths. In radio, the field has been studied at 150\,MHz \citep{Intema2011,Williams_2016,Tasse_2021}, at 325\,MHz \citep{Coppejans2015} and at 1.4\,GHz \citep{deVries2002}. The 1.4\,GHz Faint Images of the Radio Sky at Twenty Centimetres (FIRST) survey \citep{White_1997} also covers the field with a $5\,\sigma$ sensitivity limit at 0.75\,mJy.
As demonstrated in this work, the observations at 400\,MHz of the field are important in understanding source characterization, spectral studies, and source clustering etc.

The variation of the extra-galactic source counts with respect to the flux densities is directly linked to the galaxy evolution and properties \citep{Prandoni_2001,Padovani_2011,Padovani_2015}. The normalised source counts at fainter flux densities ($<1\,$mJy) involve plenty of discussion, as many deep observational studies reported a flattening of the source count distribution at around 1\,mJy \citep{Windhorst_1985, Williams_2016, Prandoni_2018}. This indicates an additional increase of radio sources at fainter flux densities, including SFGs and RQ\,AGN \citep{Jackson_1999,Rawlings_2004,Padovani_2009,Prandoni_2018}, whereas, the powerful RL\,AGN makes up the majority of the population at high flux densities.

Besides, recent studies have presented the source counts of different populations based on multi-wavelength classifications. \citet{Padovani_2016} reported the source counts in the ECDFS field at 1.4\,GHz using the VLA for SFGs, RQ\,AGN and RL\,AGN, and showed the importance of RQ\,AGN at sub-mJy levels that combined 25\,per\,cent of the sample, while SFGs and AGN contribute more or less equally to the total sample. The results by \citet{Bonato2021} confirm that SFGs and RQ\,AGN AGN dominate the counts at fainter flux density regimes, whereas the bulk of the population above $\sim300\,\upmu$Jy at 1.4\,GHz are of RL\,AGN. These outcomes are based on their study of the LH field using the WSRT observations that cover $\simeq1.4\,\rm deg^2$.

Further, in our previous work \cite{Akriti_2022}, we investigated the radio--IR relations for the ELAIS\,N1 field at 1.4\,GHz, 400\,MHz and also using the integrated radio luminosity ($L_{\rm RC}$) in the frequency range 0.1--2\,GHz. It has been found that the bolometric radio--IR relations are the strongest of all correlations with a slightly non-linear slope value of $1.07\pm0.02$ as measured for the SFGs with the variation in $q_{\rm TIR}^{RC}$ values with $z$ given as, $q_{\rm TIR}^{RC} = (2.27\pm0.03)(1+z)^{(-0.12\pm0.03)}$ up to $z=2$.
The development of bolometric relations aimed to eliminate the dependence of radio continuum measurements on the physical nature of sources. This approach has facilitated a more targeted exploration of the underlying physical parameters that can influence the evolution of radio-IR relations. The observed redshift dependence of $q$ values and the non-linearities in the slope values suggest that the evolution of magnetic fields with redshift or changes in cosmic ray acceleration efficiency may provide a plausible explanation.
To gain deeper insights into the non-linear correlation between radio and IR emissions, it is essential to acquire additional data and conduct further investigations into the evolution of magnetic fields. In this study, we utilise archival data from the uGMRT to examine radio-IR relations and their evolution in the \boot field, in a distinct region of the sky.

This paper reports the uGMRT observations of the \boot field for the first time at 400\,MHz and discuss in detail the catalogued sources, including their classification and characterization.
The paper is organised as follows. In section \ref{sec_obs}, we describe the data reduction and the imaging, followed by the source cataloguing in \ref{sec:cat}. Here, we present a comparison of our catalogue with other existing radio catalogues, followed by deriving the corrected source counts. The sources in our catalogues are also classified in SFGs, RQ\,AGN and RL\,AGN and are presented in Section \ref{sec:class_multi}. Finally, we discuss the radio--IR relations in section \ref{sec:radIR} and summarise everything in section \ref{sec:summary}.
In this work, we assume a flat cold dark matter ($\Lambda$CDM) cosmology $\Omega_\Lambda = 0.7$, $\Omega_{\rm m} = 0.3$ and $H_0 = 70\,\rm km\,s^{-1}\,Mpc^{-1}$.

\section{Observations and Data Analysis}\label{sec_obs}

We used archival data of the \boot field ($\alpha_{2000} = 14^{h}31^{m}45^{s},\, \delta_{2000} = 34^{d}28^{m}27^{s}$) observed using uGMRT in the GTAC cycle 35 (Project-code: 35\_048). These observations are centred at 400\,MHz with a total bandwidth of 200\,MHz. The total observation time is 30\,hrs ($\sim 23\,$hrs on-source) with roughly 3\,hrs for each of the seven pointing. The field was observed for a total of four days between February 3 and March 15, 2019. Table \ref{tab:obs} summarizes the observation details. The flux calibrator 3C\,286 was observed at the beginning of each session. The intermittent observation of 3C\,286 were made, which is also the phase calibrator because it lies near the target. Therefore, after every observation of two target pointings, 3C\,286 was observed. 

This data was opted for because of the lack of study of the field considered here at this frequency. This observation also covers a large area of the sky at such a low frequency. We describe the data reduction procedure and imaging in the following sections.

\begin{table}
\caption{Observation summary of the target field \boot and the calibrator sources. }
	
\begin{tabular}[width=\columnwidth]{ll}
\hline
\hline
Project code & 35\_048 \\
Observation date & 3 \& 4 Feb 2019 \\
                 & 4 \& 15 Mar 2019\\
\hline
Bandwidth &  200 MHz\\
Frequency range & 300-500 MHz\\
Channels & 8192\\
Integration time & 5.37s\\
Total on-source time & 23 h \\
Correlations & RR RL LR LL\\
\hline
Pointing centres & $13^{h}31^{m}08^{s}$  $+30^{d}30^{m}33^{s}$ (3C286)\\
                 
                 & $14^{h}31^{m}45^{s}$  $+34^{d}28^{m}27^{s}$ (\boot, 3.5 h)\\
                 & $14^{h}31^{m}45^{s}$ $+33^{d}41^{m}55^{s}$ (\boot, 3.5 h) \\
                 & $14^{h}31^{m}45^{s}$ $+35^{d}14^{m}60^{s}$ (\boot, 3.5 h) \\
                 & $14^{h}29^{m}04^{s}$ $+34^{d}51^{m}43^{s}$ (\boot, 3.3 h ) \\
                 & $14^{h}29^{m}04^{s}$ $+34^{d}05^{m}11^{s}$ (\boot, 3.3 h) \\
                 & $14^{h}34^{m}27^{s}$ $+34^{d}05^{m}11^{s}$ (\boot, 3.3 h) \\
                 & $14^{h}34^{m}27^{s}$ $+34^{d}51^{m}43^{s}$ (\boot, 2.8 h) \\
                 
\hline                 
\hline                 
\end{tabular}
\label{tab:obs}	    
\end{table}

\subsection{Data Reduction}

In our previous work on the ELAIS\,N1 field \citep{Arnab2019}, we have used a set of {\tiny CASA} based tools for Band 3 uGMRT data. These are publicly available on GitHub repository as a {\tiny CASA} based pipeline, ACAL\footnote{\href{https://github.com/Arnab-half-blood-prince/uGMRT\_Calibration\_pipeline}{https://github.com/Arnab-half-blood-prince/uGMRT\_Calibration\_pipeline}}. We have used this pipeline for radio frequency interference (RFI) mitigation and direction-independent (DI) calibrations. The DI calibration was performed on each night separately with this pipeline, and in the end, all calibrated data were combined to get the full continuum image. The following is a brief overview of the procedures involved in the pipeline.

The pipeline starts with the initial flagging of the data, which includes flagging at the edge of each scan and channel, followed by applying \texttt{TFCROP} on each of the fields. The flux density of the flux calibrator, 3C286 was then set using the model given by \citet{Perley_2017}.
The next step in the pipeline involved conducting initial delay, gain, and bandpass calibrations on the flux calibrator. This process involved selecting the reference antenna and solution interval dynamically, based on the proportion of data flagged during the generation of gain solutions. To obtain the gain solutions, the integration time is used as the solution interval. The amount of data flagged is estimated at the end, and if this is greater than 5\,per\,cent, then to obtain better solutions, a solution interval of increased value is used. These initial solutions were then applied to the primary calibrators, and flagging of the calibrated data was done using \texttt{TFCROP} and \texttt{RFLAG} to remove strong RFIs. The task \texttt{CLIP} was also used if the amplitude was greater than ten times the flux values as set by the model to flag the data per antenna. This loop of initial calibrations and flagging was repeated at least two more times, but with less strict restrictions to get rid of the bad data. Finally, a round of final calibrations was done to get the initial delay, bandpass, and gain solutions for the primary calibrator's calibrated model.

The change in gain with time was addressed by solving it with a short solution interval for the flux calibrator. Using this solution, we established the fluxscale for the entire frequency range of the primary calibrator. In the end, a final round of calibrations was run to obtain the delay, bandpass, and gain solutions, which were further transferred to the target fields. Again, \texttt{TFCROP} and \texttt{RFLAG} were used to perform flagging on the corrected data column of all the fields of the observation. Now, we split the calibrated data of all seven target pointings and proceed with the imaging and self-calibration loop.

\subsection{Imaging}

The imaging of all the pointings was done individually using \texttt{WSCLEAN} \citep{Offringa_2014}. For wide-field imaging, \texttt{WSCLEAN} uses the $w-$snapshot algorithm to correct for the array's non-coplanar nature. To capture the fluctuation of sky brightness over this wide bandwidth across several spatial scales, the multi-scale wide-band deconvolution \citep{Offringa_Smirnov_2017} is employed.
We selected a Briggs robust parameter value of $-1$ that would provide nearly uniform weighting in the data. This results in a nearly central Gaussian point spread function (PSF) and minimises broad wings.
We create a large image of approximately $3\deg^2$ in size and pixel size $1.5''$ to incorporate any bright sources outside the field of view.
A first round of images was made with 50k iterations down to 8$\,\sigma$ using the auto-masking algorithm of \texttt{WSCLEAN}. The Multi-Frequency Synthesis (MFS) image was then used to create a mask that was used for another round of imaging to fill the \texttt{MODEL\_DATA} column for performing self-calibrations. This step was performed to create a model column that is less affected by the imaging artefacts and thus, provides better solutions during the self-calibration loops.

\subsection{Self-Calibration \& Final Image}

We used \texttt{STIMELA} \citep{makhathini2018}, which is a container-based framework that consists of various radio astronomy-based software. We have used it to perform self-calibration and imaging loops for our purposes. We performed only 5 rounds of phase self-calibration loops with solution intervals as 8\,mins, 6\,mins,4\,mins,2\,mins and 1\,min. The re-imaging of the target field was done after incorporating the latest calibration solutions, and the mask was again created with the MFS image. We avoid performing amplitude or amplitude-phase self-calibrations that can affect the flux values.

\begin{figure*}
    \centering
    \includegraphics[scale=0.75]{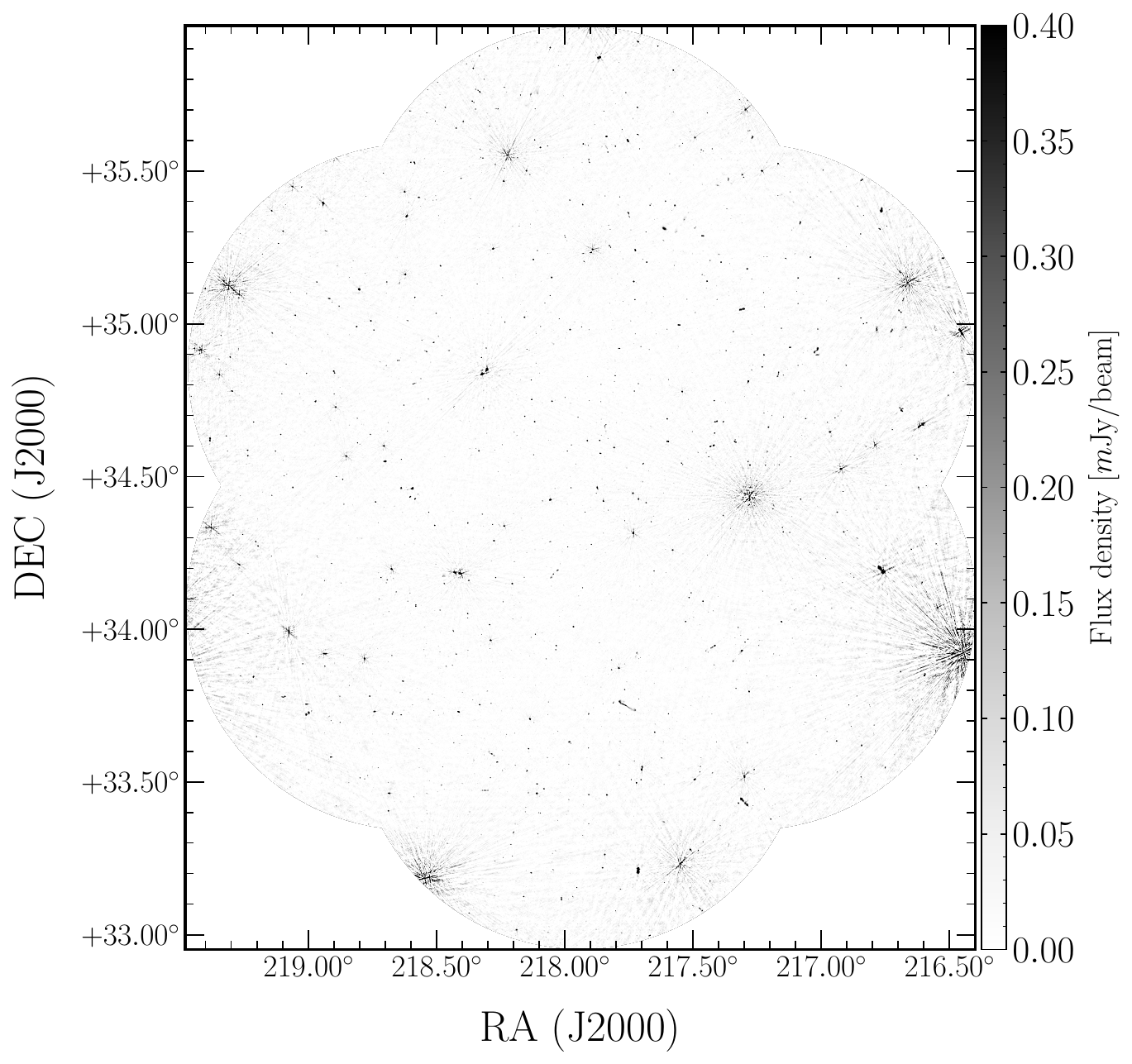}
    \caption{The primary beam corrected mosaic of \boot field at 400\,MHz using uGMRT. The central off-source RMS noise is $\sim 35\,\upmu\rm Jy\,beam^{-1}$ and the beam size is $5.0''\times4.9''$. }
    \label{fig:mosaic}
\end{figure*}

\begin{figure*}
    \centering
    \includegraphics[scale=0.5]{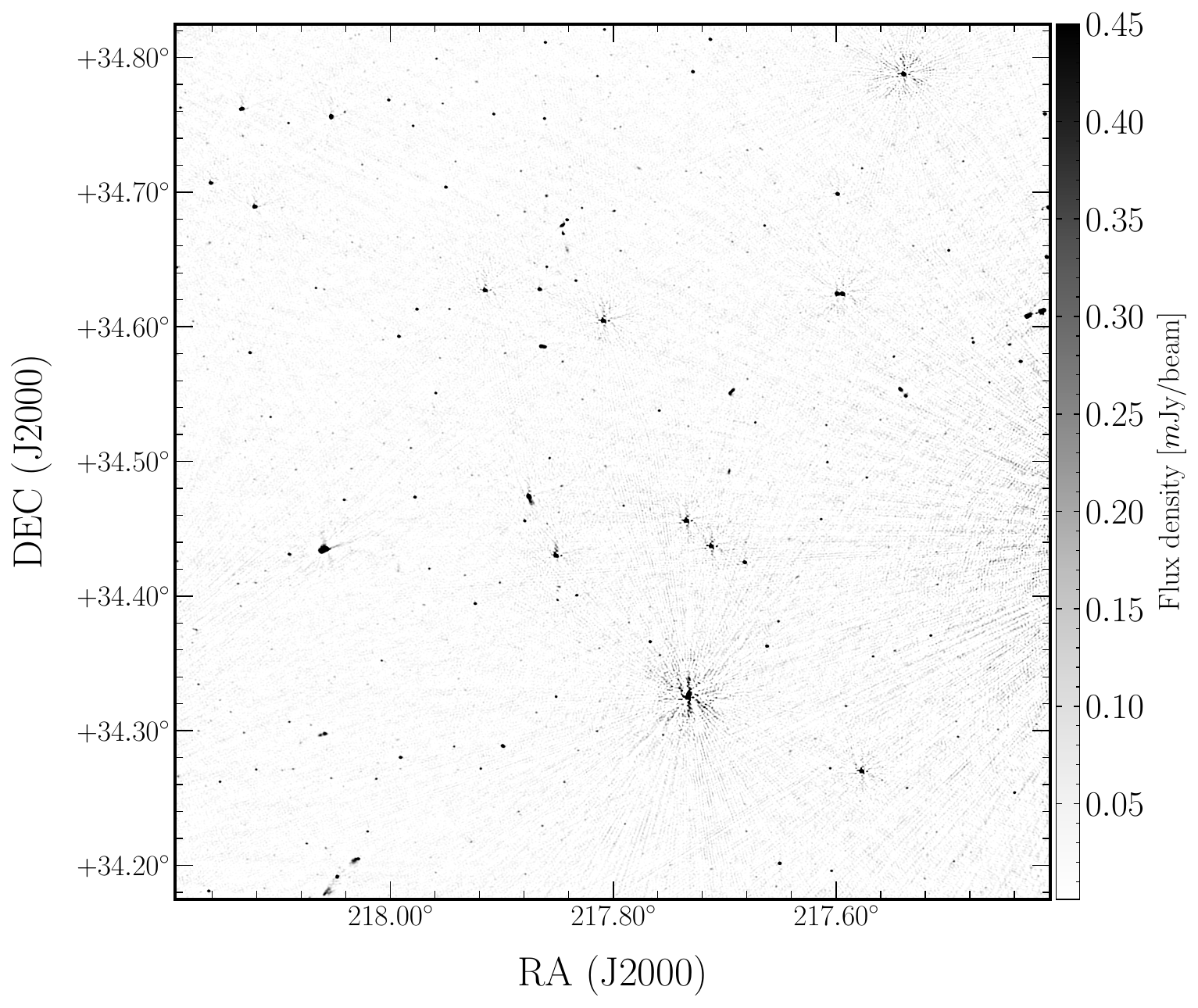}
    \caption{The total intensity image of the \boot field obtained using uGMRT at 400\,MHz covering an area of $\sim 1\deg^2$ of the central region. The off-source RMS noise measured at the centre is $\sim 35\,\upmu\rm Jy\,beam^{-1}$. }
    \label{fig:mosaic_zoom}
\end{figure*}

After processing for self-calibration and imaging all the pointings separately, each was corrected for the frequency-dependent uGMRT primary beam model\footnote{\href{http://www.ncra.tifr.res.in/ncra/gmrt/gmrt-users/observing-help/ugmrt-primary-beam-shape}{http://www.ncra.tifr.res.in/ncra/gmrt/gmrt-users/observing-help/ugmrt-primary-beam-shape}}. Here, we have imposed a cut of 20$\%$ of the primary beam response.
Thereafter, we used \texttt{MONTAGE}\footnote{\href{ http://montage.ipac.caltech.edu/}{ http://montage.ipac.caltech.edu/}} for linear mosaicking of the seven pointings to make the final combined image. In order to stitch the images together linearly with \texttt{MONTAGE}, each primary beam corrected image is first weighted by the square of the primary beam pattern, which is believed to be proportional to the noise variance image. Figure \ref{fig:mosaic} shows the final mosaic of the \boot field covering an area of $\sim6\deg^2$. We reached a minimum central off-source RMS noise of $\rm35\,\upmu Jy\,beam^{-1}$ with a beam size of $5.0''\times4.9''$. The zoomed-in central region of the image is shown in Figure \ref{fig:mosaic_zoom}.
It is observed that the direction-dependent (DD) gain calibration errors were sustained because of the bright sources in the field, as is evident from the error pattern around these sources. The DI calibration alone was not able to remove these patterns and emphasised the need for DD calibration, which will also result in improved RMS noise. A detailed analysis using direction-dependent calibrations is deferred to future work.

\section{Source Catalog }\label{sec:cat}

We have used P{\tiny Y}BDSF\footnote{For more details, please see: \href{https://www.astron.nl/citt/pybdsf/}{https://www.astron.nl/citt/pybdsf/} \label{foot1}} \citep{Mohan_Raffrey2015} to compile a source catalogue for the detection and characterization of sources. The primary beam-corrected mosaicked image was used to generate the source catalogue of the region. A sliding box window of size 170 pixels every 40 pixels, i.e., \texttt{rms\_box} = (170,40), was used to measure the varying RMS across the field. To account for the high signal-to-noise ratio around the bright artefacts and to avoid counting them as a real source, a small box \texttt{rms\_box\_bright} = (38,8) was used near them. These bright regions were first identified whose peak amplitude exceeded an adaptive threshold of 150$\sigma_{\rm RMS}$, where $\sigma_{\rm RMS}$ denotes the clipped RMS across the entire region. P{\tiny Y}BDSF locates continuous regions of emission above a pixel threshold identified as an island and models each island by fitting multiple Gaussian components.
Therefore, for source detection, we used \texttt{thresh\_isl} = $3\sigma$ and \texttt{thresh\_pix} = $5\sigma$ as the limits to include the flux for fitting in the source.

P{\tiny Y}BDSF identifies and groups adjacent Gaussians that belong to the same emission island into a single source. The source's total flux is calculated by summing the individual Gaussian fluxes, and its uncertainty is determined by combining the uncertainties of the Gaussians by adding them in quadrature.
We use the centroid of the source to determine its position and employ a moment analysis in combination with the image-restoring beam to obtain its size.

There could be some deviation in the PSF from the actual restoring beam. This issue of PSF variation across the entire map was addressed by using the parameter \texttt{psf\_vary\_do = True} in P{\tiny Y}BDSF. This selects sources with high SNR ($>20\sigma$) and that are more likely to be unresolved (here 880). More details can be found in the link to the P{\tiny Y}BDSF documentation available in the footnote \ref{foot1}.

\begin{figure*}
    \centering
    \includegraphics[width= 9cm]{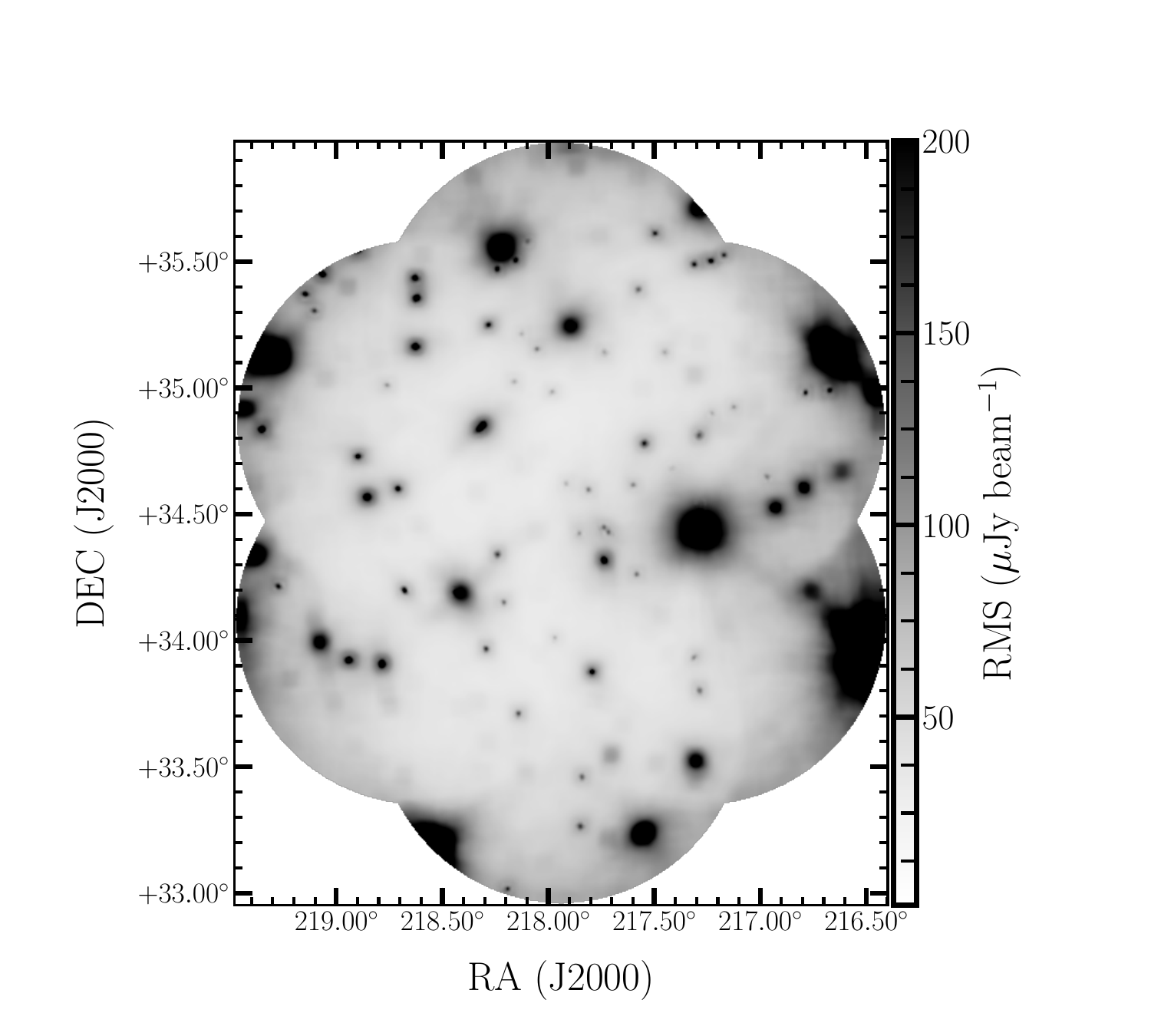}
    \includegraphics[height = 7cm]{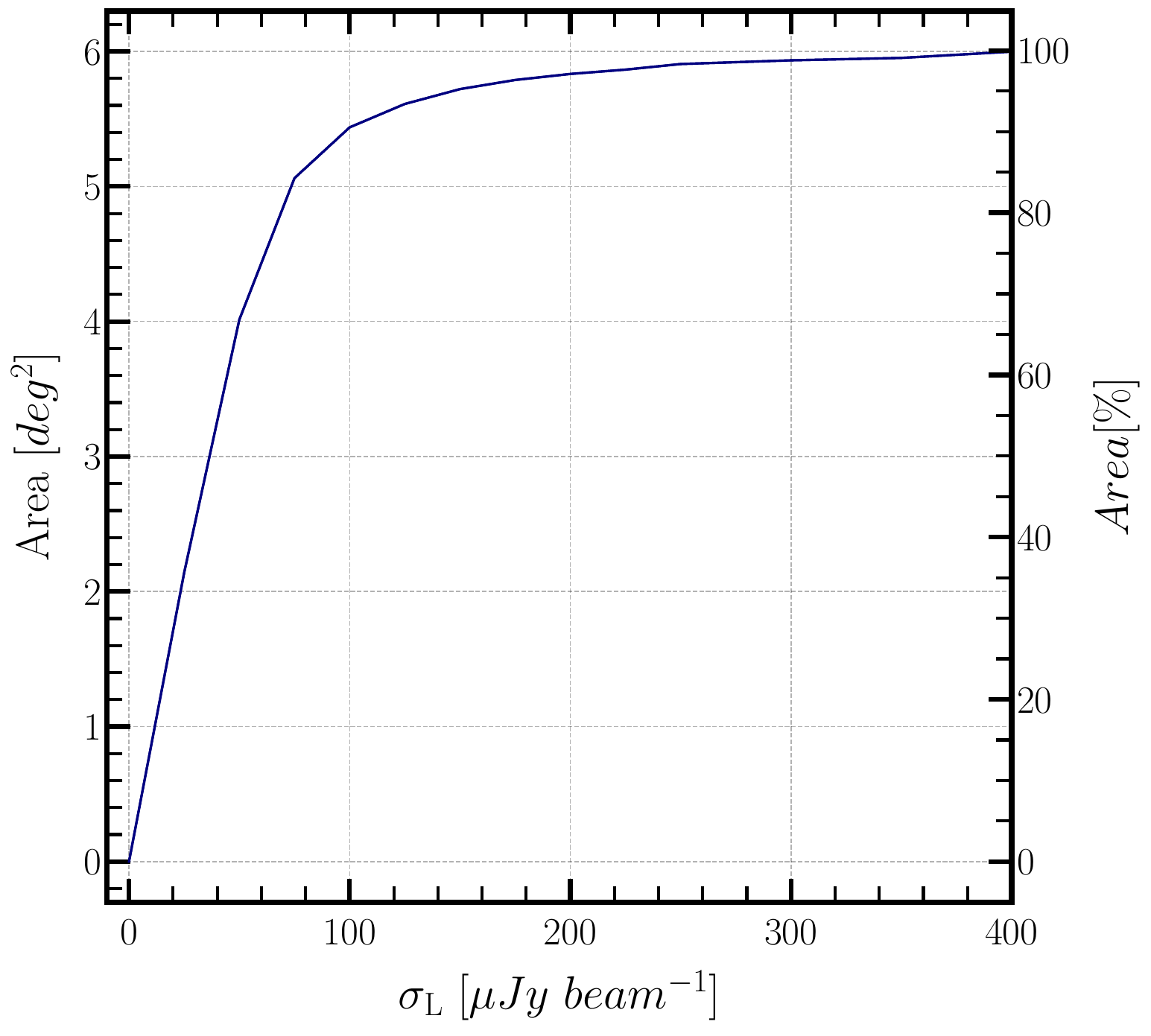}
    \caption{ \textit{Left-panel:} Variation of the local RMS noise in the final mosaicked image. \textit{Right-panel:} The cumulative area from the final mosaicked image as a function of the local RMS noise.  }
    \label{fig:rms}
\end{figure*}

P{\tiny Y}BDSF generates not only a source catalogue but also a residual map and an RMS map. The residual map is the image with all the modelled sources subtracted, and the RMS map depicts the noise variation across the field. The RMS map of the field is shown in the left-panel of Fig. \ref{fig:rms}. This shows the variation of background RMS across the field, with high values at the edge of the field and particularly near the bright sources.

\begin{table*}
\caption{Sample of the \boot field source catalogue$^\dagger$ at 400\,MHz from uGMRT.} \label{tab:cat}
\scalebox{1.0}{
\begin{tabular}{l l l l l l l l l l l l}
\hline \hline
Id & RA  & E\_RA & DEC & E\_DEC & Total\_flux &  Peak\_flux & Major & Minor & PA & rms  \\

() & (deg) & (arcsec) & (deg) & (arcsec) & (mJy)& (mJy $\mathrm{beam}^{-1}$) & (arcsec) & (arcsec) & (degree) & (mJy $\mathrm{beam}^{-1}$) \\
(1) & (2)  & (3)    & (4) & (5)     & (6) & (7) & (8) & (9) & (10) & (11)\\
 \hline
\hline
  0 & 219.4935 & 0.469 & 34.7482 & 0.321 & 0.85 & 0.582 & 6.1 & 4.9 & 86.29 & 0.091\\
  1 & 219.4913 & 0.079 & 34.7624 & 0.064 & 3.306 & 2.866 & 5.4 & 4.4 & 61.70 & 0.094\\
  2 & 219.4788 & 0.516 & 34.9168 & 0.366 & 1.607 & 1.253 & 5.7 & 4.6 & 74.29 & 0.237\\
  3 & 219.4802 & 0.147 & 34.9187 & 0.098 & 6.499 & 4.713 & 6.0 & 4.7 & 85.10 & 0.237\\
  4 & 219.4802 & 0.101 & 34.9204 & 0.114 & 7.559 & 5.446 & 5.9 & 4.8 & 36.11 & 0.237\\

\hline
\end{tabular}}
\begin{flushleft}
$^\dagger$The electronic version of the catalogue is available where the columns include the source ids, positions, flux densities and peak flux densities along with their respective errors. It also include the sizes, position angle and the local RMS noise.
\end{flushleft}
\end{table*}

We compiled a total of 3782 sources at 400\,MHz within 20\,per\,cent of uGMRT primary beam above $5\sigma$ threshold and flux densities $> 175\,\upmu$Jy. In Table \ref{tab:cat}, we show a subset of the catalogue, and the complete catalogue can be accessed through the electronic version of the paper. In the next section, we discuss the selection of the point and resolved sources.

\subsection{Compact vs Resolved sources}

\begin{figure}
    \centering
    \includegraphics[width=8cm]{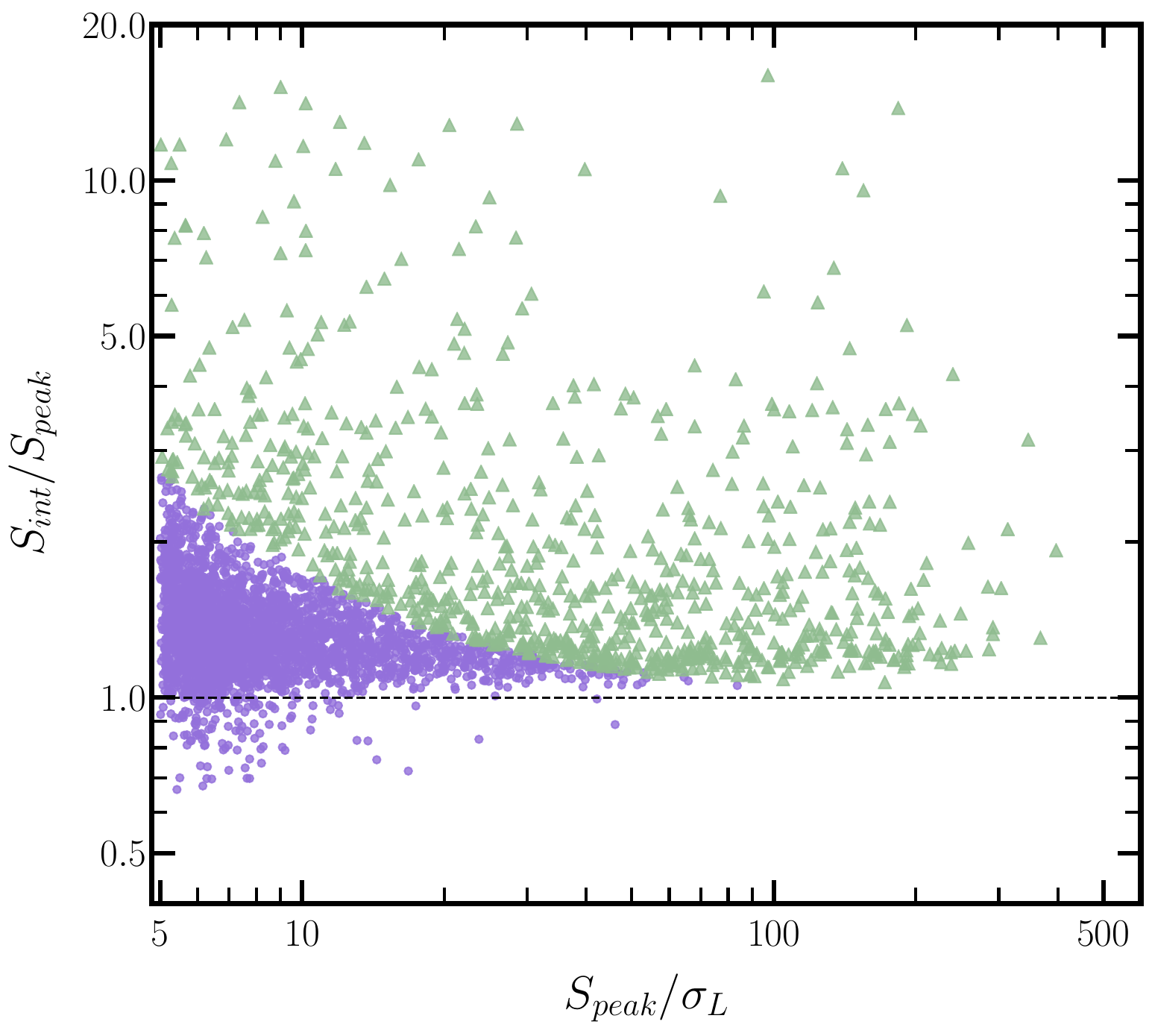}
    \caption{The variation of the ratio of the integrated to peak flux densities ($S_{\rm int}/S_{\rm peak}$) with a signal-to-noise ratio ($S_{\rm peak}/\sigma_{L}$) for the uGMRT sources. The point sources are shown in violet dots while the green triangles represent the resolved sources. }
    \label{fig:resolve}
\end{figure}

Resolved sources can be readily classified with the help of the ratio of the flux densities, i.e., integrated to the peak flux densities, $\Sint/\Speak> 1$, for noise-free scenarios. However, uncertainties due to calibration and variable noise may skew the $\Sint/\Speak$ distribution. This skewing effect is evident, especially at low SNR, as shown in Fig. \ref{fig:resolve}, which displays the variation of $\Sint/\Speak$ with $\Speak/\sigma_{\rm L}$, where $\sigma_{\rm L}$ is the local RMS. Based on the source parameters derived from P{\tiny Y}BDSF, it is difficult to identify whether the detected sources are point-like or resolved. To some extent, the effect of time and bandwidth smearing can falsely extend sources in the image plane.

The classification of resolved and point sources is performed following the methods outlined in \cite{Franzen_2015,Franzen_2019}. Assuming that the uncertainties in the peak flux density ($\sigma_{S_{\rm peak}}$) and integrated flux density ($\sigma_S$) are independent of each other, $ln(S/S_{\rm peak})$ obeys a Gaussian distribution with a mean centred around zero. The RMS would then be given by:
\begin{equation}
    \sigma_{\rm R} = \sqrt{ \Big(\frac{\sigma_S}{S_{\rm int}} \Big)^2 + \Big(\frac{\sigma_{S_{\rm peak}}}{S_{\rm peak}}\Big)^2}
\end{equation}

Therefore, a source is identified as extended at the 3$\sigma$ level if $ln(S/S_{\rm peak}) > 3\sigma_{\rm R}$ \citep{Franzen_2015}. In our sample, we classify 794 sources as resolved and 2988 sources as point sources shown in the green triangles and violet dots, respectively, in Figure \ref{fig:resolve}.

\begin{table}
    \centering
    \caption{Details of the catalogues considered. The columns represent the catalogue, frequency in MHz, resolution, corresponding RMS noise and their $5\sigma$ sensitivity in mJy. The last column present the matched sources after applying the selection criteria as discussed in Section \ref{sec:flux}.} 
    \begin{tabular}{cccccc}
     \hline \hline
    Catalogue & Frequency & Resolution & $\sigma$  & $S_{\rm limit}$ & Matched*  \\
        & (MHz) & (arcsec) & (mJy) & (mJy)& sources  \\
    \hline

    uGMRT & 400 & $5.0''$ & 0.035 & 0.175 &  \\
    FIRST & 1400 & $5.4''$ & 0.20 & 1.0 & 210 \\
    WSRT & 1400 & $27''$ & 0.028 & 0.14 & 469 \\
    VLA  & 325 & $5.6''$ & 0.20$^\dagger$ & 1.0 & 526 \\
    LOFAR (16) & 150 & $7.4''$ & 0.12 & 0.60 & 722 \\
    LOFAR (21) & 150 & $6''$ & 0.03 & 0.15 & 761 \\
    \hline
    \end{tabular}
    \begin{flushleft}
        \footnotesize{\textit{Note:} The values in column four are the RMS noises measured across the field except for $^\dagger$ that gives the central RMS noise. * are the number of sources after applying the selection criteria in section \ref{sec:flux}.}
    \end{flushleft}
    
    \label{tab:survey}
\end{table}

\subsection{Comparison With Other Radio Catalogs}

Comparing with other radio catalogues of the same region is crucial, considering the uncertainties in the uGMRT primary beam model and the significant ionospheric fluctuations at low frequencies. Because of these reasons, the source may get smeared at a large distance from the phase centre with distorted source positions. Therefore, systematic differences in flux density and source positions can be quantified by utilizing the multi-frequency catalogues available in the literature. In this section, we compare our catalogue to other radio catalogues that have overlapping coverage of the same sky area. We have used the FIRST survey and Westerbork Synthesis Radio Telescope (WSRT) observations of the \boot field by \citet{deVries2002} for comparison at 1.4\,GHz. In addition at low frequencies, we have utilized LOFAR observations at 150\,MHz by \citet{Williams_2016} and \citet{Tasse_2021}, and, VLA observations at 325\,MHz by \citet{Coppejans2015}. 

We have identified the counterparts of our uGMRT sources in other catalogues by using a search radius of $3''$. Based on observation sensitivity and completeness, each catalogue will have its flux density limits ($S_{\rm limit}$).  
For our analysis here, we have selected those sources only whose flux densities are higher than $S_{\rm limit}$. The details of the various catalogues used are presented in Table \ref{tab:survey}.

\subsubsection{Flux Accuracy}\label{sec:flux}

In this section, we compare flux densities of the \boot field at 400\,MHz with previous radio catalogues. 
We have used \citet{Perley_2017} flux density scales. 
The low-frequency observations by \citet{Williams_2016} and \citet{Tasse_2021} at 150\,MHz has used \citet{Scaife2012} flux scale. At higher frequency, we selected the FIRST catalogue at 1.4\,GHz for comparison where \citet{Baars1977} scale was used. We have also compared with 325\,MHz from \citet{Coppejans2015} from the VLA and at 1.4\,GHz from \citet{deVries2002} from the WSRT observations. 

We follow \citet{Williams_2016} to select the sample for comparison: (i) the sources with a high signal-to-noise ratio ($> 10$), (ii) compact sources whose sizes are less than the higher resolution catalogue, and (iii) isolated sources where the minimum distance between two sources is greater than twice the PSF for a catalogue with lower resolution. We have made sure to convert the flux density values of the catalogues to the scale proposed by \cite{Perley_2017}, wherever available.

\begin{figure}
    \centering
    \includegraphics[width = 8cm]{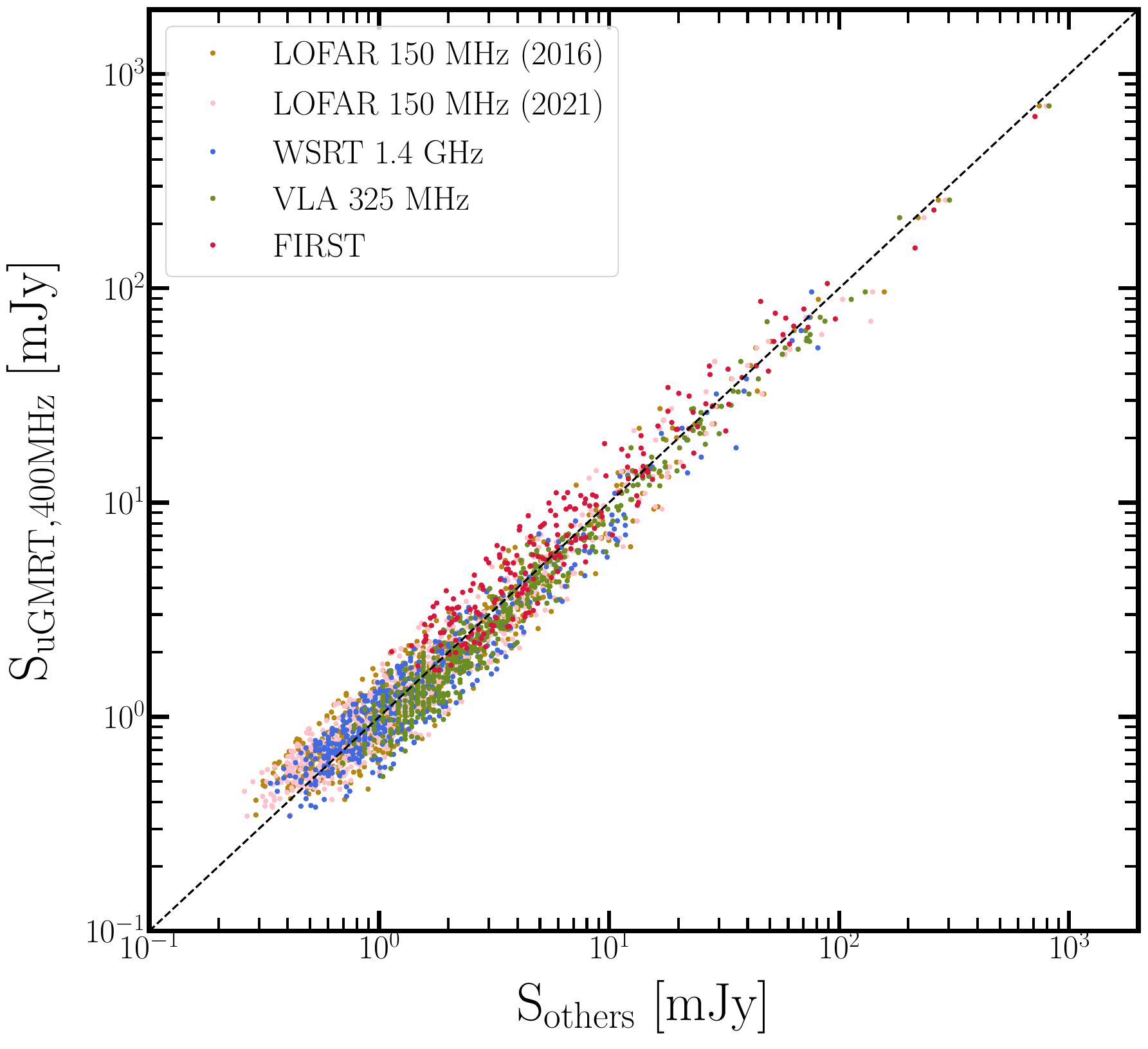}
    \caption{The variation of total integrated flux densities at 400\,MHz from uGMRT observations for compact sources with radio catalogues at various frequencies. For this comparison, we have used 150\,MHz LOFAR (yellow and pink), 325\,MHz VLA (green) and 1.4\,GHz from WSRT (blue) and FIRST (red), respectively. $S_{\rm uGMRT}/S_{\rm other}=1$ is shown by the dashed black line.  }
    \label{fig:flux_comp}
\end{figure}

We derive the ratio between the total flux density at 400\,MHz with other catalogues scaled to 400\,MHz using $\alpha=-0.7$, defined as $S_{\rm 400\,MHz}/S_{\rm others}$. The comparison of $S_{\rm 400\,MHz}$ with $S_{\rm others}$ is also shown in Fig. \ref{fig:flux_comp} and we do not find any critical deviation from the $S_{\rm 400\,MHz}/S_{\rm other}=1$ line shown in black dashed line. The median $S_{\rm 400\,MHz}/S_{\rm others}$ ratio derived using the LOFAR catalogues as reported by \cite{Williams_2016} and \cite{Tasse_2021} are $ 0.98_{-0.2}^{+0.3}$ and $0.98_{-0.2}^{+0.3}$, respectively. Here, the errors are from the 16th and 84th percentiles. The median values of this ratio measured using the WSRT, FIRST and VLA are $0.98_{-0.3}^{+0.3}, 0.88^{+0.2}_{-0.3}$, and $0.87^{+0.2}_{-0.1}$, respectively. The ratios are nearly 1 for almost all the cases suggesting negligible systematic offsets. However, we found a median ratio value of around 0.88 when compared to the FIRST catalogue, 
which could have been raised due to flux calibration error, given that the error estimates provided in the catalogues are limited to fitting errors. It is also possible that further errors may arise due to the assumption of a constant spectral index for each source.

\subsubsection{Spectral Index Distribution}

\begin{figure}
    \centering
    \includegraphics[width = 8cm]{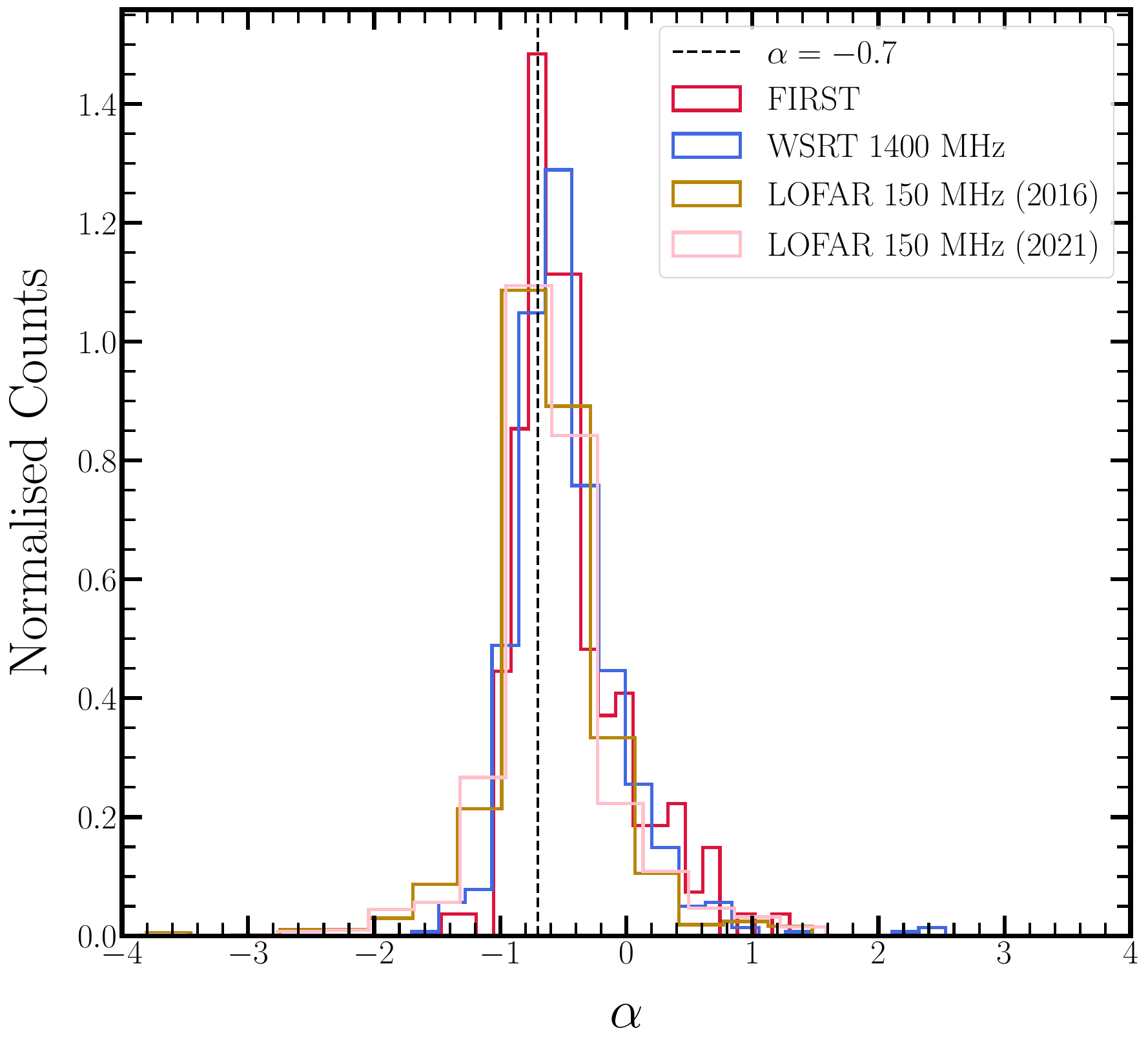}
    \caption{The normalised spectral index distribution measured in comparison with the other radio catalogues. We have used LOFAR at a lower frequency and WSRT and FIRST at higher frequencies to measure the spectral indices. The dashed line is at $\alpha=-0.7$ and is shown for reference.  }
    \label{fig:alpha}
\end{figure}

We assume a synchrotron power-law distribution ($S\propto\nu^\alpha$) to determine the spectral index distribution. We estimated the two-point spectral indices using the FIRST catalogue at 1.4\,GHz, WSRT catalogue at 1.4\,GHz \citep{deVries2002} and the two LOFAR catalogues at 150\,MHz \citep{Williams_2016,Tasse_2021}. 
The matched sources (following the selection criteria in the previous section) in the uGMRT and other catalogues are presented in Table \ref{tab:survey}.
Figure \ref{fig:alpha} shows the normalised histogram of the spectral indices estimated from the above-mentioned catalogues.

The median spectral value with errors from the 16th and 84th percentile using various catalogues are: $-0.63^{+0.3}_{-0.3}$ (150\,MHz LOFAR; \citealt{Williams_2016}), $-0.63^{+0.3}_{-0.2}$ (150\,MHz LOFAR; \citealt{Tasse_2021}), $-0.69^{+0.3}_{-0.2}$(1.4\,GHz WSRT) and $-0.61^{+0.2}_{-0.2}$ (1.4\,GHz FIRST).
The median spectral indices measured using the WSRT catalogue are consistent, but for other cases, we observe slightly flat values of median spectral indices in comparison to \citet{Coppejans2015}, who found median $\alpha =-0.72$ between 325--1400\,MHz using the FIRST catalogue. Also, the study by \cite{Williams_2016} reported a median value of $-0.79$ between 150--1400\,MHz. 
Nevertheless, considering the errors, our measures of median spectral indices values are within the uncertainties. 
It should be noted that the calibration errors and sample selection could potentially introduce a bias.
Furthermore, we performed an analysis to investigate any potential biases that may arise due to the flux density limits of different surveys at various frequencies. We confirm that there are no such biases involved in this study. A detailed investigation of spectral indices of the sources in the field, including direction-dependent calibration and in-band $\alpha$ measurements, is deferred to future work.

\subsubsection{Positional Accuracy}

\begin{figure}
    \centering
    \includegraphics[width = \columnwidth]{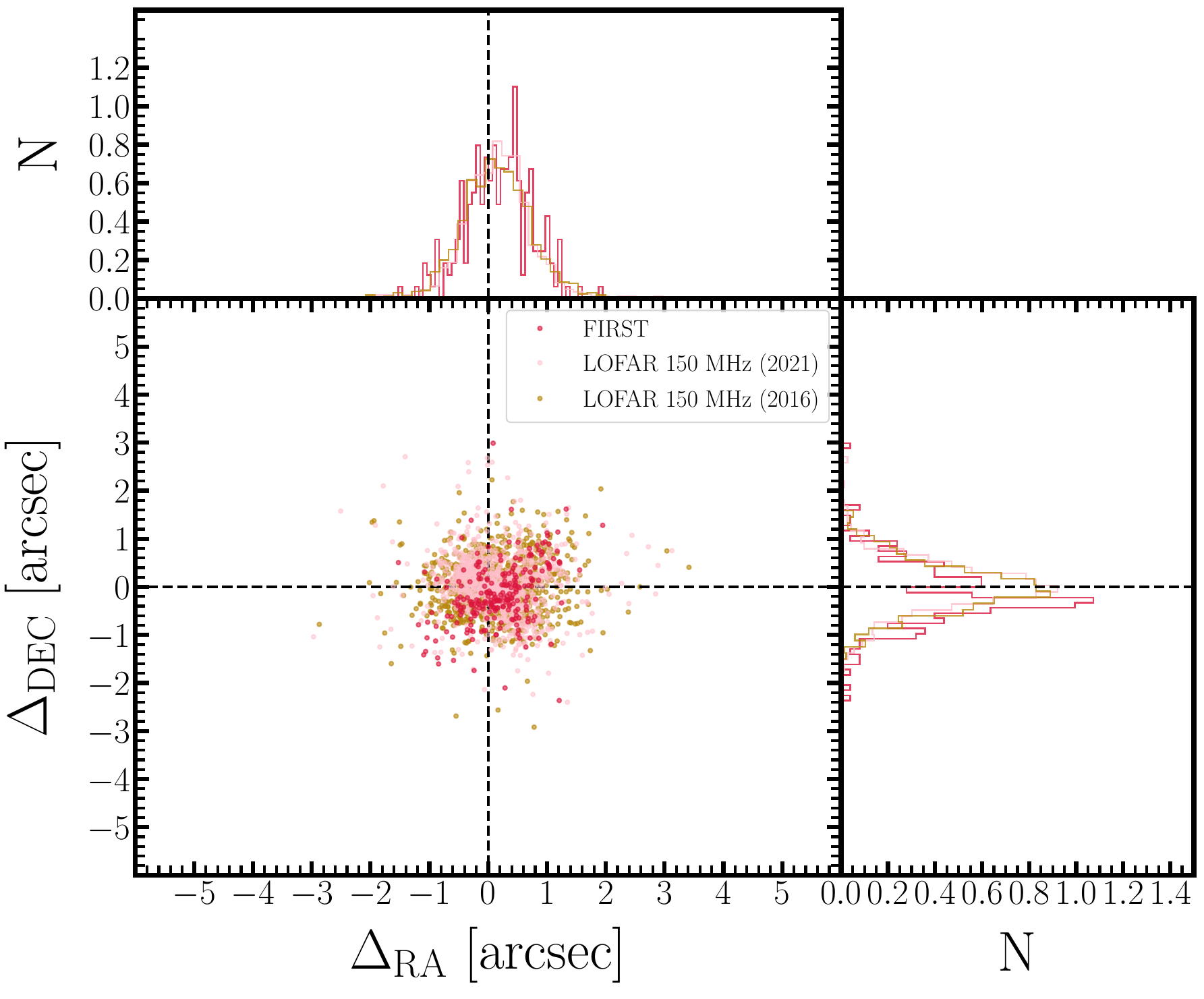}
    \caption{The positional offset in RA and DEC for the uGMRT sources at 400\,MHz from the FIRST (red), LOFAR 150\,MHz \citep[pink;][]{Tasse_2021} and LOFAR 150\, MHz \citep[yellow;][]{Williams_2016} catalogues.}
    \label{fig:offset}
\end{figure}

The positional accuracy for our uGMRT sample is obtained by comparing their positions with the source positions from the FIRST, VLA 325\,MHz, LOFAR 150\,MHz (2016), LOFAR 150\,MHz (2021) catalogues. The resolution of these catalogues is comparable to our resolutions of $5.0''\times4.9''$. The criterion to select a source sample for this study is the same as mentioned in section \ref{sec:flux}. Among all catalogues, FIRST \citep{Becker1995} is at a higher frequency (1.4\,GHz) having less ionospheric fluctuations and has a fairly better resolution of $5.4''$ \citep{White_1997} with positional accuracy better than $1''$. We follow \citet{Williams_2016} to measure the offsets in right ascension (RA) and declination (DEC) for our uGMRT sample at 400\,MHz as:
\begin{align}
\begin{split}
    \delta_{\rm RA} &=  \rm{RA}_{\rm uGMRT}  - \rm{RA}_{\rm FIRST} \\
    \delta_{\rm DEC} &=  \rm{DEC}_{\rm uGMRT} - \rm{DEC}_{\rm FIRST}
\end{split}
\end{align}
Table \ref{tab:offsets} shows the median offset values estimated from the various catalogues used. The median values of the deviation in RA and DEC measured using the FIRST catalogue are $0.204''$ and $-0.255''$. In Fig. \ref{fig:offset}, the offsets in RA and DEC for uGMRT source catalogue from other catalogues are shown along with their histograms. The positional offsets don't show systematic variation across the field of view. Moreover, the offsets measured from the LOFAR catalogues (see Table \ref{tab:offsets}) are negligible when compared to the uGMRT image cell size of $1.5''$. We have applied a correction on source positions using the constant median offset values estimated based on the FIRST catalogue to our final uGMRT catalogue.

\begin{table}
    \centering
        \caption{The median values of the offsets in RA and DEC measured in our uGMRT source catalogue when compared to other catalogues.}
    \label{tab:offsets}
    \begin{tabular}{cccc}
    \hline \hline
    Catalogue & Frequency & $\delta_{\rm RA,median}$ & $\delta_{\rm DEC,median}$\\
    \hline 
        FIRST & 1400 &  0.204 & -0.255 \\
        LOFAR (2016) & 150 & 0.149 & -0.024 \\ 
        LOFAR (2021) & 150 & 0.163 & 0.005 \\
\hline

    \end{tabular}

\end{table}

\subsection{Source Counts}

In this section, we use the uGMRT catalogue to compute the differential source counts at 400\,MHz. The results from observations and simulations \citep{Bonaldi_2018,Wilman_2008} have found that SFGs and radio-quiet quasars (RQQs) are the populations that predominate at faint flux densities. However, at sub-mJy levels ( primarily $<$0.5\,mJy), the detected source population is also subjected to relatively few observational limitations. In fact, fainter sources ($\leq \upmu$Jy) act as viable foregrounds for faint HI 21\,cm cosmological signal.
The spatial and spectral characterization of sources is important to understand the nature of the Universe at large.

Here, we have estimated the differential source count down to 100\,$\upmu$Jy (>\,5$\sigma$) at 400\,MHz using the uGMRT catalogue. However, the source count directly measured using the P{\tiny Y}BDSF's output catalogue may not portray the true distribution and could be biased because of various errors like false detection, catalogue incompleteness, resolution and Eddington bias. This becomes challenging, particularly at lower frequencies and when dealing with sources at the lower limits of the flux density distribution. Therefore, the corrections on the source counts must be applied to account for these biases to study the true nature of the extra-galactic source distribution. These are described and discussed in the following subsections.

\begin{figure}
    \centering
    \includegraphics[width = 7cm]{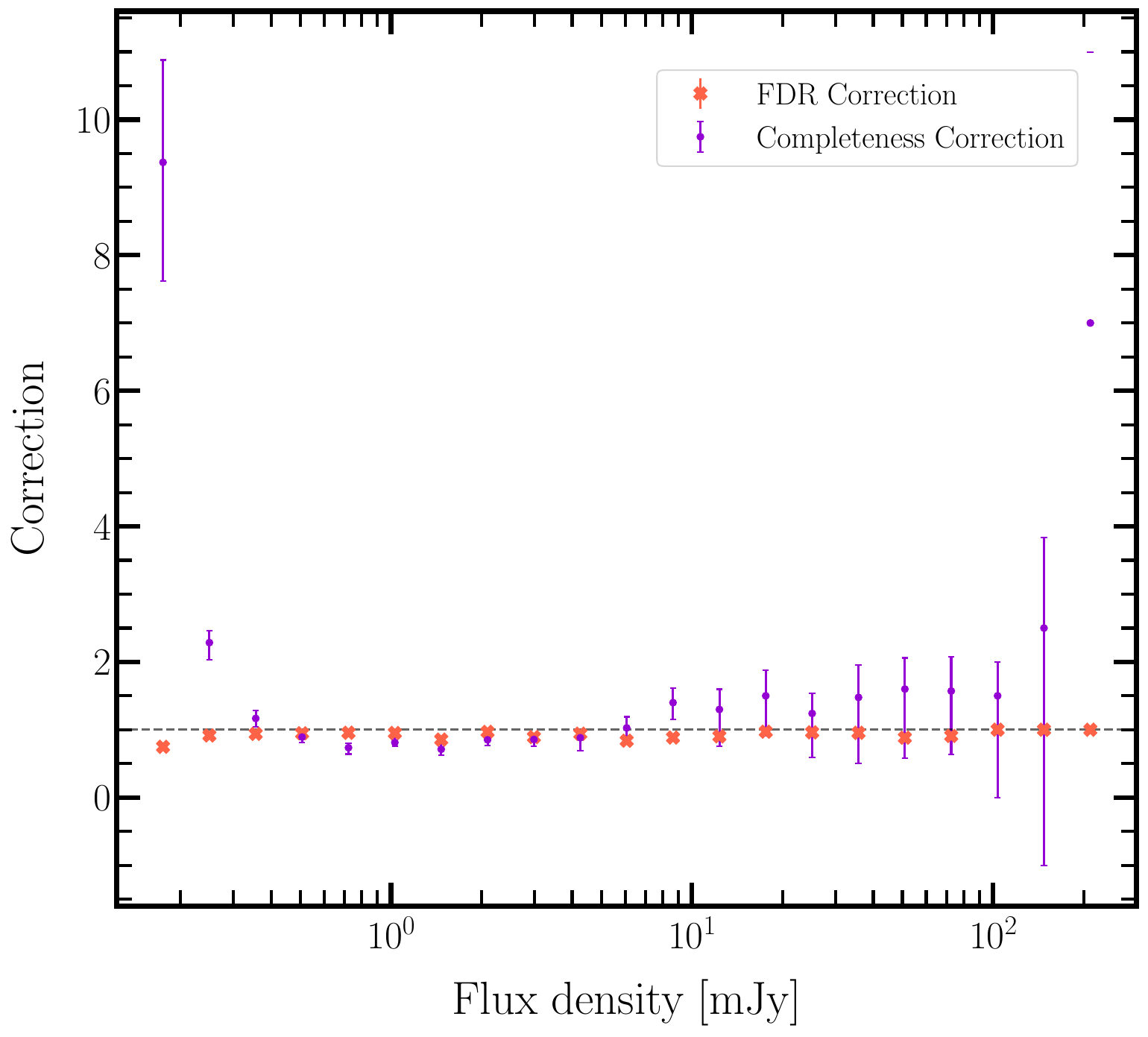}
    \caption{The correction factors due to FDR and completeness are shown in orange and violet respectively. }
    \label{fig:corr}
\end{figure}

\begin{figure*}
    \centering
    \includegraphics[scale=0.45]{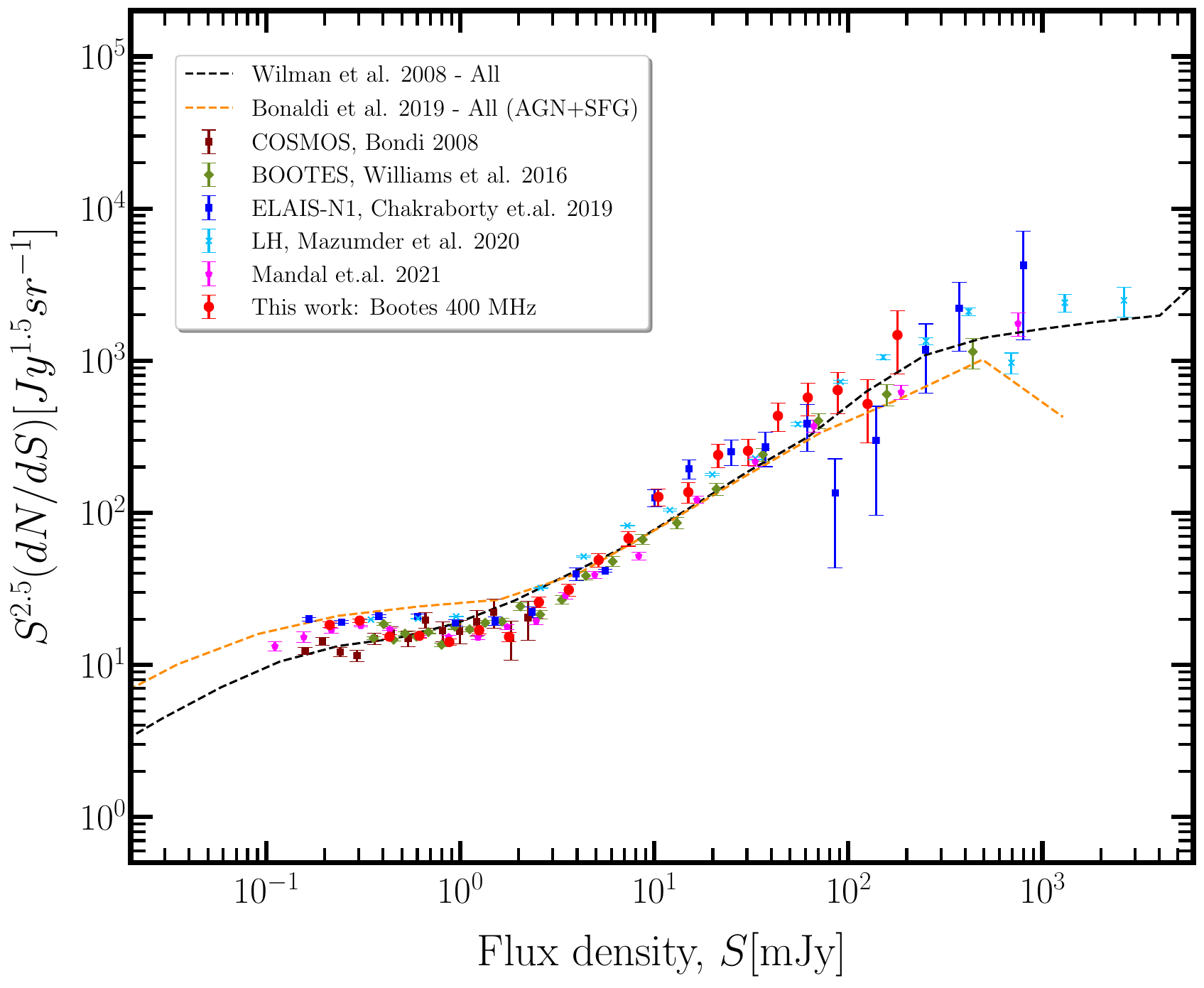}
    \caption{The Euclidean normalised differential source counts for the 400\,MHz uGMRT observations shown in red circles after applying the correction factors including false detection and incompleteness. The source counts of the same field at 150\,MHz from \citet{Williams_2016} (in green diamonds) and \citet{Mandal_2021} (in pink left-triangles), and of other fields, such as the COSMOS \citep{Bondi2008} in brown squares, the ELAIS-N1 field \citep{Arnab2019} in blue squares and the LH field \citep{Mazumder2020} in cyan crosses are shown for comparison. The black dashed and yellow dashed line represents the source counts modelled by $S^3-$SKADS \citep{Wilman_2008} and T-RECS \citep{Bonaldi_2018} simulations, respectively.  }
    \label{fig:sc_all}
\end{figure*}

\subsubsection{FDR}

The number of fake sources that P{\tiny Y}BDSF mistakenly identified as real ones is known as the false detection rate (FDR). This may occur because of the noise spikes or bright artefacts in the image. The number of negative sources detected in the inverted (negative) image will represent total false detections assuming noise distribution to be symmetric around zero, i.e., positive noise peaks have equal negative peaks in the image.
To address this issue, we run P{\tiny Y}BDSF on the negative image with parameters similar to those used for the original image (see section \ref{sec:cat}). Here, we have found that the RMS map obtained from the inverted image does not represent the RMS map obtained from the real image. Therefore, we use RMS map of the original image to finally compile the catalogue from the inverted image. With negative peaks less than $-5\sigma$, we have found 288 sources in total. 

To correct for FDR in the flux density bins, negative sources from the inverted image were binned similarly to real sources and compared to the positive sources obtained from the original image.
Following \citet{Hale2019}, the proportion of actual sources in every bin would then be:
\begin{equation}
    f_{\rm{real},i} = \frac{N_{\rm{catalog},i} - N_{\rm{inv},i}}{N_{\rm{catalog},i}}
\end{equation}
where $N_{\rm{catalog},i}$ are the number of sources detected in $i^{\rm th}$ flux density bin from the original image and $N_{\rm{inv},i}$ denotes the same but from the inverted image. The errors on FDR are estimated using Poissonian errors. This fraction is multiplied to the corresponding number of sources in each flux bin from the original catalogue. Fig. \ref{fig:corr} shows the FDR correction factor measured for each bin in orange crosses.

\begin{table*}
\caption{Estimates of the Euclidean normalised differential source counts for the \boot field. The columns represent the flux density bins, the central value of the flux density bins, and the normalised differential source counts ($S^{2.5}\rm{d}N/\rm{d}S$) in units of $\rm{Jy}^{1.5}\rm{sr}^{-1}$, FDR, completeness and corrected normalised source counts.} \label{tab:sc_all}
\begin{tabular}{|r|r|r|r|r|r|r|}
\hline \hline
  \multicolumn{1}{|c|}{$S$} &
  \multicolumn{1}{c|}{$S_{\rm{c}}$} &
  \multicolumn{1}{c|}{N} &
  \multicolumn{1}{c|}{$ S^{2.5}$dN/dS} &
  \multicolumn{1}{c|}{FDR} &
  \multicolumn{1}{c|}{Completeness} &
  \multicolumn{1}{c|}{Corrected $ S^{2.5}$dN/dS}  \\
  (mJy) & (mJy) &  & (Jy$^{1.5}\rm{sr^{-1}}$) &  &  &  (Jy$^ {1.5}\rm{sr^{-1}}$) \\
\hline
  0.15-0.22 & 0.21 & 111 & 2.61 $\pm$ 0.11 & 0.65 $\pm$ 0.01 & 9.37$_{-1.75}^{1.51}$ & 18.33 $\pm$ 0.79\\
    & & & & & &  \\
  0.22-0.31 & 0.30 & 558 & 9.30 $\pm$ 0.28 & 0.98 $\pm$ 0.0 & 2.28$_{-0.25}^{0.18}$ & 19.54 $\pm$ 0.58\\
      & & & & & &  \\
  0.31-0.44 & 0.43 & 741 & 13.99 $\pm$ 0.44 & 0.87 $\pm$ 0.0 & 1.17$_{-0.12}^{0.11}$ & 15.35 $\pm$ 0.48\\
      & & & & & &  \\
  0.44-0.64 & 0.61 & 686 & 18.34 $\pm$ 0.66 & 0.84 $\pm$ 0.0 & 0.89$_{-0.09}^{0.08}$ & 15.54 $\pm$ 0.56\\
      & & & & & &  \\
  0.64-0.92 & 0.88 & 470 & 20.06 $\pm$ 0.90 & 0.81 $\pm$ 0.0 & 0.74$_{-0.09}^{0.07}$ & 14.15 $\pm$ 0.64\\
      & & & & & &  \\
  0.92-1.32 & 1.25 & 310 & 21.90 $\pm$ 1.23 & 0.70 $\pm$ 0.01 & 0.81$_{-0.06}^{0.10}$ & 16.89 $\pm$ 0.095\\
        & & & & & &  \\
  1.32-1.89 & 1.78 & 213 & 25.27 $\pm$ 1.72 & 0.69 $\pm$ 0.01 & 0.71$_{-0.08}^{0.08}$ & 15.30 $\pm$ 1.04\\
        & & & & & &  \\
  1.89-2.71 & 2.54 & 156 & 31.20 $\pm$ 2.50 & 0.83 $\pm$ 0.01 & 0.85$_{-0.08}^{0.12}$ & 25.84 $\pm$ 2.06\\
        & & & & & &  \\
  2.71-3.90 & 3.62 & 120 & 40.85 $\pm$ 3.72 & 0.79 $\pm$ 0.01 & 0.86$_{-0.10}^{0.09}$ & 31.15 $\pm$ 2.84\\
        & & & & & &  \\
  3.90-5.60 & 5.16 & 102 & 59.00 $\pm$ 5.84 & 0.87 $\pm$ 0.01 & 0.88$_{-0.20}^{0.10}$ & 48.92 $\pm$ 4.84\\
        & & & & & &  \\
  5.60-8.04 & 7.36 & 80 & 78.71 $\pm$ 8.80 & 0.82 $\pm$ 0.01 & 1.03$_{-0.24}^{0.16}$ & 67.90 $\pm$ 7.59\\
        & & & & & &  \\
  8.04-11.55 & 10.49 & 61 & 102.13 $\pm$ 13.07 & 0.87 $\pm$ 0.01 & 1.40$_{-0.25}^{0.21}$ & 127.25 $\pm$ 16.29\\
        & & & & & &  \\
  11.55-16.58 & 14.96 & 41 & 116.82 $\pm$ 18.24 & 0.90 $\pm$ 0.01 & 1.30$_{-0.53}^{0.30}$ & 136.27 $\pm$ 21.33\\
        & & & & & &  \\
  16.58-23.81 & 21.32 & 34 & 164.89 $\pm$ 28.28 & 0.85 $\pm$ 0.01 & 1.50$_{-0.58}^{0.38}$ & 239.91 $\pm$ 41.14\\
        & & & & & &  \\
  23.81-34.20 & 30.40 & 26 & 214.62 $\pm$ 42.09 & 0.85 $\pm$ 0.01 & 1.24$_{-0.65}^{0.30}$ & 255.48 $\pm$ 50.10\\
        & & & & & &  \\
  34.20-49.11 & 43.34 & 22 & 309.11 $\pm$ 65.90 & 0.86 $\pm$ 0.01 & 1.48$_{-0.98}^{1.48}$ & 433.73 $\pm$ 92.47\\
          & & & & & &  \\
  49.11-70.53 & 61.79 & 17 & 406.58 $\pm$ 98.61 & 1.00 $\pm$ 0.00 & 1.60$_{-1.02}^{-0.46}$ & 572.46 $\pm$ 138.84\\   
         & & & & & &  \\
  70.53-101.29 & 88.08 & 11 & 447.81 $\pm$ 135.02 & 1.00 $\pm$ 0.00 & 1.57$_{-0.94}^{0.51}$ & 640.19 $\pm$ 193.03\\
         & & & & & &  \\
  101.29-145.46 & 125.57 & 5 & 346.48 $\pm$ 154.95 & 0.80 $\pm$ 0.04 & 1.50$_{-1.50}^{0.50}$ & 519.72 $\pm$ 232.43\\
        & & & & & &  \\
  145.46-208.90 & 179.02 & 5 & 589.77 $\pm$ 263.75 & 1.00 $\pm$ 0.00 & 2.50$_{-3.50}^{1.34}$ & 1474.43 $\pm$ 659.38\\
          & & & & & &  \\

\hline
\end{tabular}
\end{table*}

\subsubsection{Completeness}\label{sec:comp}

The source catalogue generated using P{\tiny Y}BDSF may not be complete due to variation of noise across the image. Therefore, incompleteness refers to the inability of the source finder algorithm to identify sources above the given flux density threshold of the catalogue. This can cause underestimation or overestimation of the source counts measured in each flux density bin.

Several factors contribute to the underestimation of source counts, particularly in the vicinity of the flux density detection limit. Even though the flux density errors of the stars fit the typical Gaussian distribution, \citet{Eddington1913} demonstrated a considerable bias in the calculated star counts. 
The observed source count appears to be steeper because of confusion noise or system noise, or both, which can lead to an overestimation of counts, especially near the detection limit of the catalogue. 
Besides, resolution bias also affects the catalogue's incompleteness. This bias arises because of the lower probability of detecting a resolved source compared to one that appears point-like. The peak flux density for an extended source may be so low that it is not distinguishable from background noise, thus resulting in reduced source counts.

To consider a correction for the catalogue's completeness, we performed a simulation where we injected 2000 sources into our residual RMS map \citep[similar to][]{Williams_2016} randomly across the field using an open-source software Aegean \citep{Hancock2012,hancock2018}. Approximately 21\,per\,cent of these sources are simulated to have major and minor axes greater than the resolution of about $\simeq5''$, indicating that they are extended sources, while the remaining 79\,per\,cent are point sources.
The flux densities for these simulated sources are randomly drawn in the range 140\,$\upmu$Jy to 300\,mJy from a power-law distribution of the form dN/dS $ \propto S^{-1.6}$ \citep{Intema2011,Williams2013}. We performed 100 realizations of these simulations. 
The simulations also inherently consider the visibility area impacts \citep{Williams_2016,Hale2019,Franzen_2019}. Now again, we have used the same parameters as discussed in section \ref{sec:cat} to extract the sources from each simulated image using P{\tiny Y}BDSF. These extracted sources were then binned the same as the actual sources to estimate the completeness correction factor for $i^{\rm th}$ bin as \citep[following][]{Hale2019}:
\begin{equation}
    \rm{Correction}_{,i} = \frac{N_{\rm{injected},i}}{N_{\rm{recovered},i}}
    \label{eq:comp}
\end{equation}
where $N_{\rm{injected},i}$ represents the number of artificially injected sources and $N_{\rm{recovered},i}$ indicates the total number of sources identified from the simulated image. This method acquired to quantify completeness also considers the resolution and Eddington biases. 

In Fig. \ref{fig:corr}, we show the correction factor for completeness along with the FDR corrections for each flux density bin. The correction factor here for each flux bin is determined as the median value of the 100 simulations and the errors associated are from the 16th and 85th percentiles.

\begin{figure*}
    \centering
    \includegraphics[width= 5cm,height=5cm]{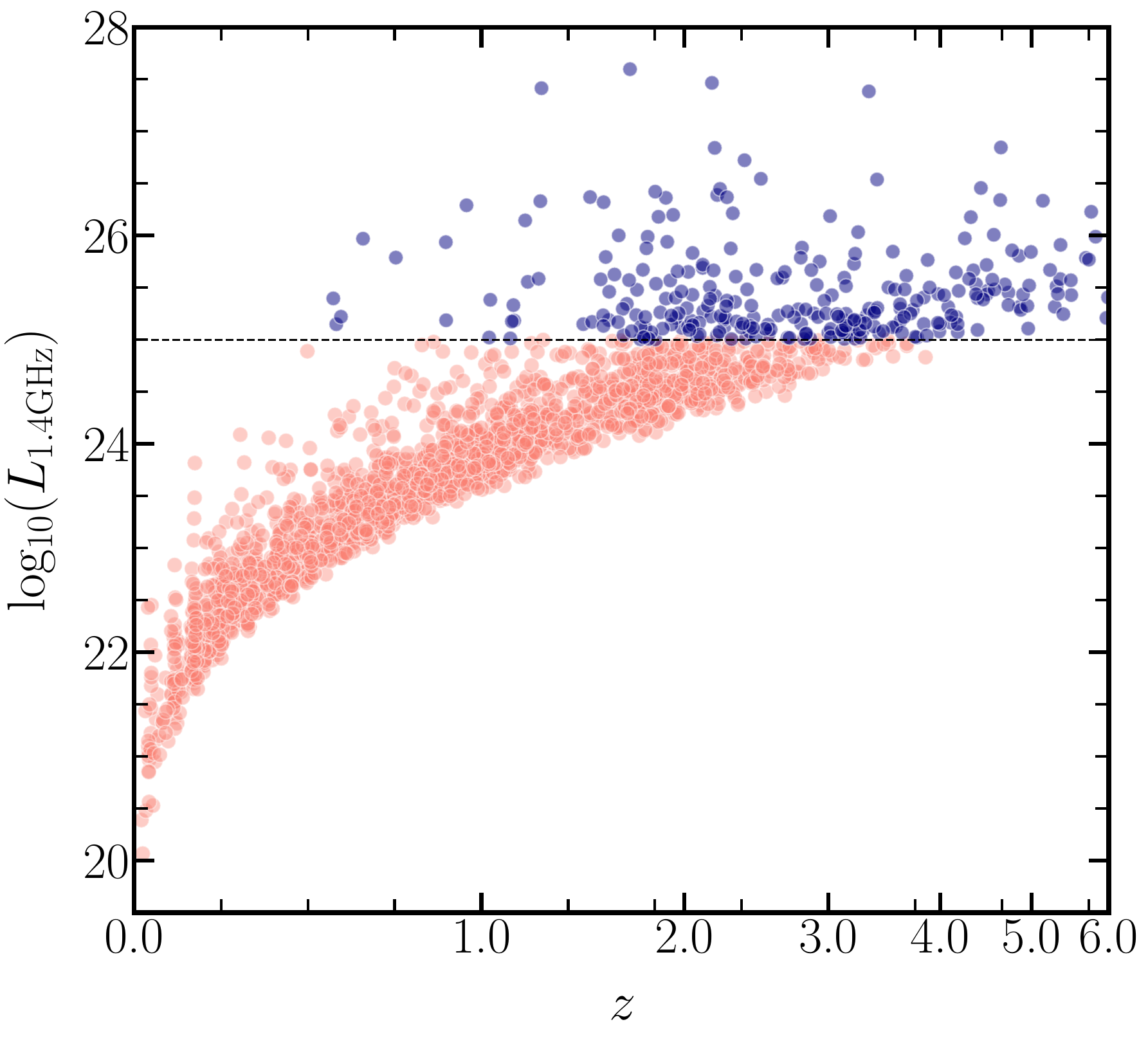}
    \includegraphics[width= 5cm,height=4.9cm]{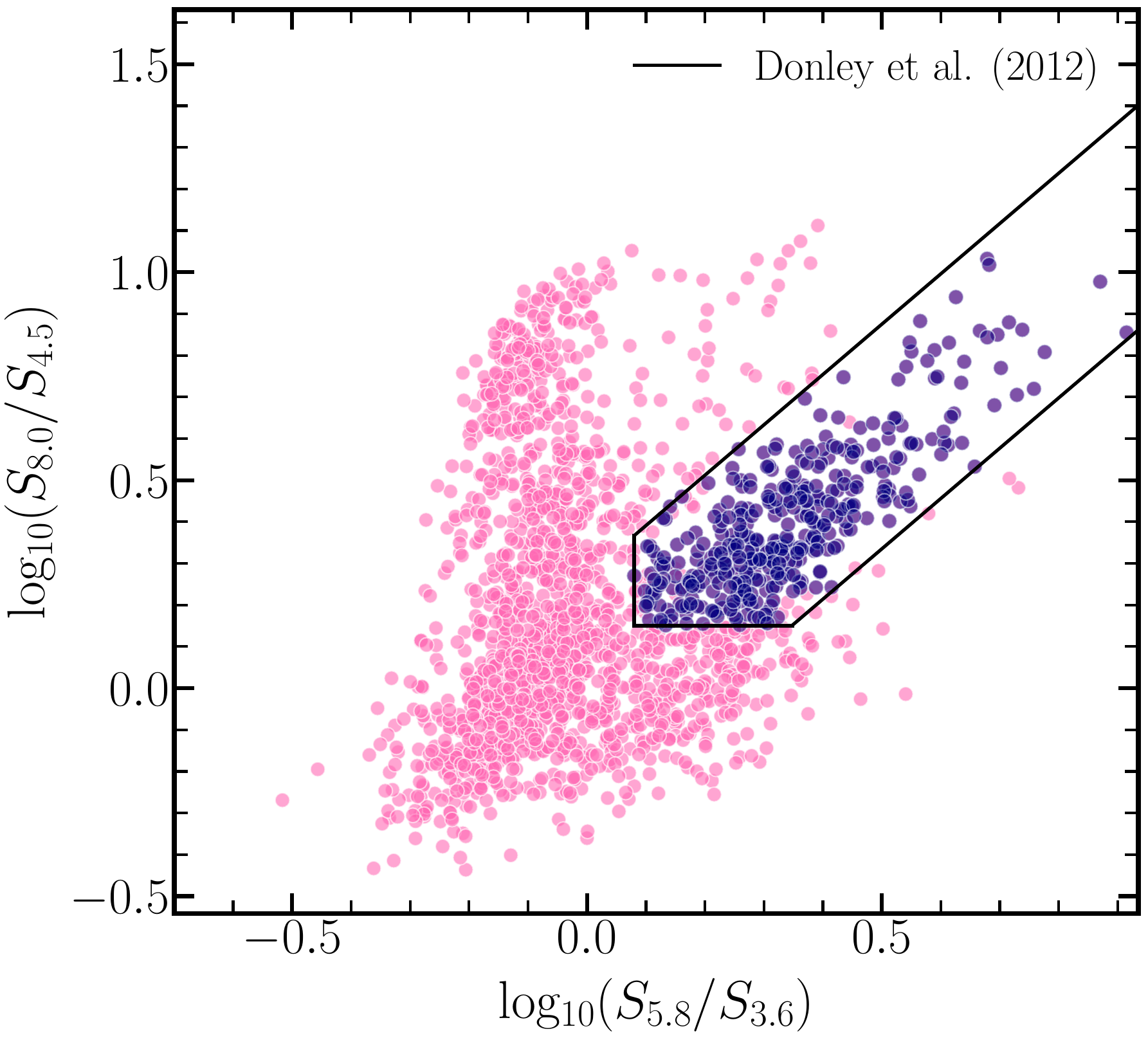}
    \includegraphics[width= 6cm,height= 4.9cm]{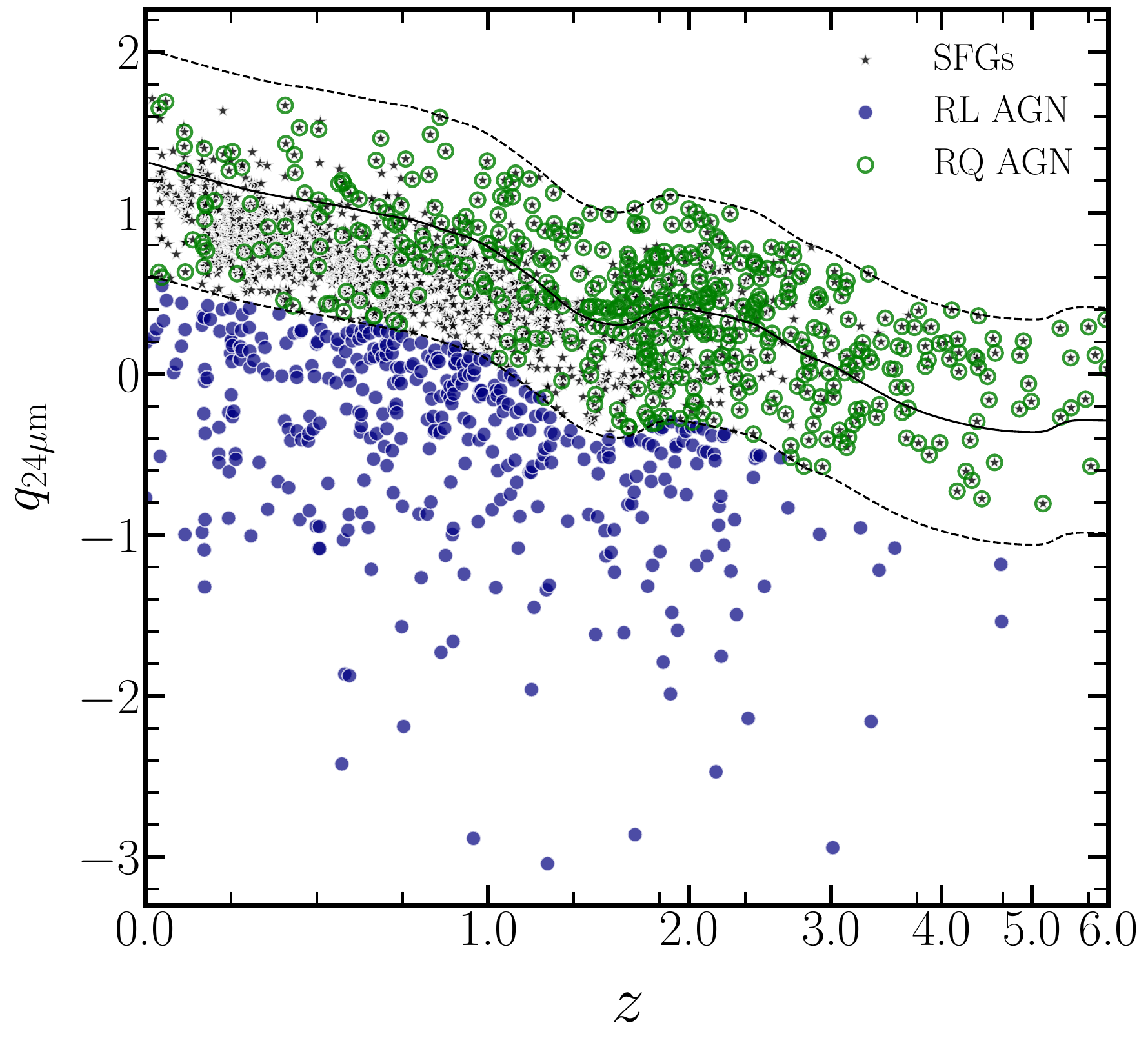}
    \caption{\textit{Left-panel:} The variation of radio luminosity ($L_{\rm 1.4\,GHz}$) with redshift ($z$). The navy-colored sources above the dashed line indicate RL AGN classification based on their luminosity of $10^{25}\rm WHz^{-1}$. \textit{Middle-panel:} The figure shows an IRAC colour-colour diagram used to identify AGN, with the solid lines representing the boundaries of the region identified by \citet{Donley_2012}. The navy blue points represent sources classified as AGN lying within this region. \textit{Right-panel:} The variation of observed $q_{\rm 24\,\upmu m}$ with $z$. The black solid line indicates the redshifted $q_{\rm 24,\upmu m}$ values of the M\,82 galaxy template for local starburst galaxies. The dashed black curves represent the dispersion at $2\sigma$ levels.
        } 
    \label{fig:class_multi}
\end{figure*}

\subsubsection{Differential Source Count-Comparison}

We derived the Euclidean-normalised differential source counts from the uGMRT catalogue at 400\,MHz. These counts have been corrected for FDR and completeness in each flux bin by multiplying the corresponding correction factors to the uncorrected source counts. Besides, Fig. \ref{fig:rms} shows the noise varies across the image. Hence, a correction is needed for the effective area per bin over which a source can be identified. The completeness simulations performed in section \ref{sec:comp} also accounts for this correction.

In Fig. \ref{fig:sc_all}, we present the normalised source counts in red points after applying the corrections discussed in the previous sections. We have used 21 logarithmically-spaced bins based on flux density, extending down to 175\,$\upmu$Jy ($5\sigma$). We estimated the Poisson errors on source count in each bin. Table \ref{tab:sc_all} shows the source counts and corresponding errors. To provide context and comparison, we display the source counts obtained from different observations and recent simulations.
We have used the differential source counts of the same field at $150\,$MHz from \citet{Williams_2016} and from \citet{Mandal_2021} that gives the combined source counts values for the \boot, ELAIS N1 and the Lockman Hole field obtained using LOFAR data. Besides, we also have utilised source counts from \citet{Bondi2008} at 1.4\,GHz, from \citet{Arnab2019} at 400 MHz, and also from \citet{Mazumder2020} at 325\,MHz in the fields, COSMOS, ELAIS-N1 and Lockman Hole, respectively. All the catalogues used are scaled to 400\,MHz using a spectral index of -0.7. 
The $S^3-$SKADS simulations \citep{Wilman_2008} and T-RECS simulations at 1.4\,GHz are also used for comparison. The source counts from these simulations were scaled similarly using $\alpha=-0.7$ to 400\,MHz. SKADS-simulation outputs a synthetic radio catalogue utilising various multifrequency observations to simulate the luminosity function, source clustering, source classification, etc. Also, AGN and SFGs, together with their respective sub-populations, were individually modelled using updated evolutionary models in the recent T-RECS simulations and the resulting output was compared with the existing data for validation \citep[see][]{Bonaldi_2018}. 

Our measurements of source counts are in agreement with both observations and simulations (see Figure \ref{fig:sc_all}). We decided to skip the last bin at the high end of the flux bins because of the exceptionally high value of source counts obtained from only two sources in that bin. From the mentioned simulations, it has been found that the source counts flatten below 1\,mJy as a result of increased faint radio sources like SFGs and RQQ at these fluxes. At higher flux densities, these are in good agreement with the previous findings and sky models. 
The increase in SFGs and RQ AGN population could be a possible reason for the increase of source counts at sub-mJy levels. Therefore, in the next  section, we classify the sources to derive the source counts for the different populations for confirmation.

\section{AGN/SFG Classification}

The \boot field is an extensively studied field with an ancillary of multi-wavelength data that would help us to classify a source in a radio catalogue as AGN or SFG. It is, however challenging in some sources to identify whether the radio emissions originate from star-forming regions in the host galaxies or due to relativistic jets powered by AGN. 
This section begins with a description of the multi-wavelength data and redshift catalogues used, followed by the classification schemes of \citet{Bonzini_2013} for identifying AGN and SFGs \citep[similar to][]{Akriti_2022} from our radio catalogue.

\subsection{Ancillary Multiwavelength Data and Redshifts}

We have used the merged spectroscopic redshift ($z_{\rm spec}$) catalogues available from the \textit{Spitzer} Extragalactic Representative Volume Survey (SERVS) Data Fusion\footnote{\href{http://www.mattiavaccari.net/df/pw/}{http://www.mattiavaccari.net/df/pw/}} \citep{Vaccari2010,Vaccari2015} that covers $11.10\,\deg^2$ of the field. This catalogue contains $z_{\rm spec}$ information from the AGN and Galaxy Evolution Survey \citep[AGES;][]{Kochanek_2012} and from the 16th Data Release of the Sloan Digital Sky Surveys (SDSS) \citep{Ahumdada_2020} in addition to the $z_{\rm spec}$ available from the NED database. About 29.3\,per\,cent (1109) of sources from the uGMRT 400 MHz catalogue have been identified in this catalogue with $z_{\rm spec}$ information. A search radius of 3 arcsecs was used to identify a cross-match in each catalogue. For the remaining sources, we have used the redshift catalogue provided by LoTSS \citep{Duncan2021} that covers $25\,\deg^2$ combining three fields including this and found 1991 sources (52.6\,per\,cent) with redshift information. Of these, 40 sources have $z_{\rm spec}$ from multiple spectroscopic data, and 1951 sources are assigned with photometric redshifts ($z_{\rm ph}$). Further, 210 additional sources also have photometric redshifts from SERVS Data Fusion \citep{Vaccari2010,Vaccari2015}. Therefore, combining all the redshift information we have 3310 sources ($\sim$87.5\,per\,cent) with either of the redshifts in our uGMRT catalogue.

\begin{table}
    \centering
        \caption{The uGMRT 400\,MHz catalogue cross-matching statistics using the multiwavelength catalogues.  }
    \label{tab:cat_multi}
    \begin{tabular}{ccc}
    \hline \hline
        Catalogue & Size & Percentage  \\
        \hline
        uGMRT 400\,MHz & 3782 & 100  \\
        Merged $z_{\rm spec}$  & 1109 & 29.3\\
        LoTSS $z$ & 1991 &  52.6\\
        $z_{\rm phot}$ from Data Fusion & 3031 &  80.1  \\
        Redshifts ($z$) & 3310 & 87.5  \\
        SERVS   & 3222 & 85.2  \\
        SWIRE IRAC all bands & 2513 & 66.4 \\
        MIPS 24\,$\upmu$m & 2262 & 59.8 \\
        
        \hline
    \end{tabular}

\end{table}

The SERVS DR1 also compiles a catalogued output from Infrared Array Camera (IRAC; at 3.6, 4.5, 5.8, 8.0\,$\upmu$m) and Multi-Band Imaging Photometer (MIPS; 24\,$\upmu$m) using \textit{Spitzer} \citep{Surace_2005,Ashby_2009,Vaccari2010,Vaccari2015}. Table \ref{tab:cat_multi} summarizes the redshift and multiwavelength catalogue used with their fraction of radio counterparts. We have used these data to identify AGN from our radio catalogue and is described in the subsequent section. We utilized the data from the Herschel Multitiered Extra-galactic Survey (HerMES) carried out by the Herschel Space Telescope, which observed at 250, 350, and 500\,$\upmu$m mapping an area of approximately 380 square degrees using the Herschel-Spectral and Photometric Imaging Receiver \citep[SPIRE;][]{Roseboom_2010,Roseboom_2012}.

\begin{figure}
    \centering
    \includegraphics[width = 8cm]{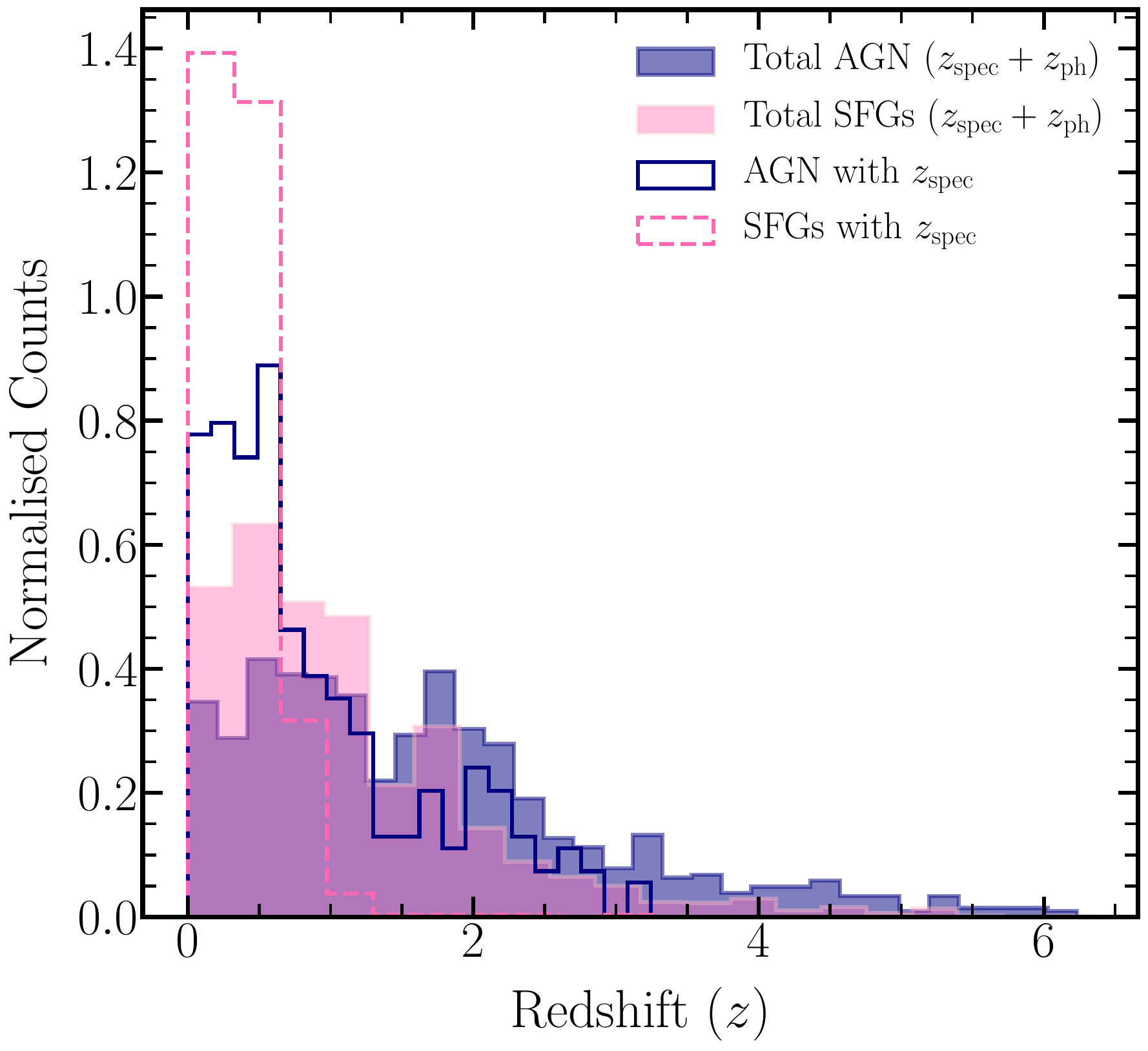}
    \caption{The redshift distribution for SFGs and AGN in the \boot field. The combined distribution for the spectroscopic ($z_{\rm spec}$) and photometric ($z_{\rm ph}$) redshifts for SFGs and AGN are shown in the filled pink and navy histograms, respectively. The dashed pink and solid navy open histograms show the $z_{\rm spec}$ distribution for SFGs and AGN, respectively.   }
    \label{fig:zhist}
\end{figure}

\subsection{Classification Synopsis}\label{sec:class_multi}

\begin{table}
    \centering
    \caption{The total unique number of SFGs, RL AGN and RQ AGN classified from the selection criteria in section \ref{sec:class_multi}. Here, the fraction is with respect to the total number of sources with redshift measurements, i.e., 3310 sources.  } \label{tab:class_multi}
    \begin{tabular}{ccccc}
         \hline \hline
         Class & & Number & Fraction (per\,cent)  \\
         \hline
         SFGs & & 2326 & 70\\
         AGN  & & 984 & 30 \\
             & RL AGN & 418 & 12.6 \\
             & RQ AGN & 496 & 15\\
        \hline
    \end{tabular}
      
\end{table}

We identify AGN from our uGMRT sample that has redshift measurements (3310), on the basis of radio luminosity, IRAC colours, spectroscopy and the ratio of the observed flux densities at 24$\upmu$m and 1.4\,GHz ($q$ values) along with the classifications obtained from the LoTSS catalogue. Each of these criteria is summarized below: 

\begin{enumerate}
    \item Radio Luminosity: The sources whose rest-frame radio luminosity at 1.4\,GHz, $L_{\rm 1.4\,GHz} > 10^{25}\,\rm WHz^{-1}$ are identified as RL\,AGN \citep{Jiang_2007,Sajina_2007,Sajina_2008}. We measured $L_{\nu}$ as,
    \begin{equation}
      L_{\nu} = 4\,\uppi\,d_{\rm L}^{2} \frac{S_{\nu}} {(1+z)^{1+\alpha}},\, \, \rm{where,} \, S_{(\nu = 1.4\,\rm GHz)} = \Big(\frac{1.4}{0.4}\Big)^\alpha S_{0.4\,\rm GHz}
    \label{eq:lum}
    \end{equation}
    Here, $d_{\rm L}$ is the luminosity distance and $\alpha$ is the spectral index ($S\propto \nu^\alpha$). We assumed a spectral index of $\alpha = -0.7$ \citep{Ibar_2010}. In the left-panel of Figure \ref{fig:class_multi}, we show the $L_{\rm 1.4\,GHz}$ variation with $z$ and classify 273 sources as RL AGN (navy).
    
    \item IRAC colours : Using the wedge suggested by \citet{Donley_2012} in the IRAC colour-colour diagram shown in Figure \ref{fig:class_multi} (middle-panel) for our uGMRT sample, we identify 381 sources as AGN (navy).
    
    \item SDSS : The merged redshift catalogue that we use also consists of the spectroscopic classification of sources from SDSS for 336 sources. The catalogue's \texttt{CLASS} and \texttt{SUBCLASS} keywords were utilised to identify 47 AGN in our radio catalogue.    
    \item LOFAR : Additionally, sources are categorised in the photometric redshift catalogue \citep{Duncan2021} given by the LOFAR using various multi-wavelength data from the literature. This collection includes AGN found using the HMQ Half Million Quasars \citep{Alam_2015} catalogue, X-ray catalogue from \citep{Kenter2005} and also using mid-IR SED criteria proposed by \citet{Donley_2012}. The flag \texttt{AGN} was used from the LOFAR redshift catalogue in order to select 504 AGN in our uGMRT sample.
    \item $q_{24\,\upmu \mathrm{m}}$ values: We measured the logarithmic ratio $q_{24\,\upmu\mathrm{m}}$ of observed mid-infrared flux at $24\,\upmu$m to radio flux density at $1.4\,\mathrm{GHz}$,
    $q_{24\, \upmu \mathrm{m}} = \mathrm{log}_{10}(S_{24\, \upmu \mathrm{m}}/ S_{\rm1.4\,GHz})$ for the sources having $24\,\upmu$m detections and redshifts to select RL AGN. The median $q_{24\,\upmu \mathrm{m}}$ value measured for the full sample is $0.49\pm0.31$ at a median $z$ of 0.88. We used the redshifted $q_{24\,\upmu \mathrm{m}}$ values of the local starburst M\,82 galaxy template that is normalised to the local average $q_{24\,\upmu \mathrm{m}}$ value \citep{Sargent_2010_2} and is shown in the Figure \ref{fig:class_multi} (right-panel) as the solid curve with a $2\sigma$ scatter represented by the black dashed line,  where $\sigma = 0.35$. The sources for which $q_{24\,\upmu \mathrm{m}}$ values lie below the $2\sigma$ template of the M\,82 curves are classified as RL AGN. These are shown in navy points in the right-panel of Figure \ref{fig:class_multi}, and we selected 387\,RL\,AGN using this criterion.
    
\end{enumerate}

\begin{table*}
\caption{Estimates of the Euclidean normalised differential source counts for the classified sources in the \boot field at 400\,MHz. The columns represent the central value of the flux density bins for all these classified sources i.e., for SFGs, RQ\,AGN and RL\,AGN along with their corresponding normalised differential source counts ($S^{2.5}\rm{d}N/\rm{d}S$ in units of $\rm{Jy}^{1.5}\rm{sr}^{-1}$.} \label{tab:sc_class}
\begin{tabular}{|r|r|r|r|r|r|r|r|}
\hline \hline
  \multicolumn{1}{|c|}{$S_{\rm SFGs}$} &
  \multicolumn{1}{c|}{Counts$_{\rm SFGs}$} &
    \multicolumn{1}{c|}{} &

  \multicolumn{1}{|c|}{$S_{\rm RQ\,AGN}$} &
  \multicolumn{1}{c|}{Counts$_{\rm RQ\,AGN}$} &
      \multicolumn{1}{c|}{} &

  \multicolumn{1}{c|}{$S_{\rm RL\,AGN}$} &
  \multicolumn{1}{c|}{Counts$_{\rm RL\,AGN}$} \\
  (mJy) & (Jy$^{1.5}\rm{sr^{-1}}$) & & (mJy) & (Jy$^{1.5}\rm{sr^{-1}}$) &  & (mJy) & (Jy$^{1.5}\rm{sr^{-1}}$)   \\
\hline
  0.30 & 18.31 $\pm$ 0.45 & & 0.25 & 5.06 $\pm$ 0.37 & & 0.42 & 1.83 $\pm$ 0.13\\
  0.65 & 10.02 $\pm$ 0.30 & & 0.47 & 1.68 $\pm$ 0.10 & & 1.32 & 5.81 $\pm$ 0.45\\
  1.41 & 4.10 $\pm$ 0.22 & & 0.87 & 0.81 $\pm$ 0.07 & & 4.16 & 18.83 $\pm$ 1.97\\
  3.06 & 2.65 $\pm$ 0.23 & & 1.61 & 0.55 $\pm$ 0.06 & & 13.16 & 42.62 $\pm$ 7.65\\
  6.63 & 3.39 $\pm$ 0.35 & & 3.0 & 0.29 $\pm$ 0.06 & & 41.62 & 139.09 $\pm$ 32.78\\
  14.37 & 6.15 $\pm$ 0.81 & & 5.58 & 0.10 $\pm$ 0.04 & & 131.62 & 200.94 $\pm$ 71.04\\

\hline
\end{tabular}
\end{table*}

Combining all the above criteria and counting unique sources, we classify 984 sources as AGN from our uGMRT sample. 
For our analysis, we classify the sources that have redshift measurements and were not identified as AGN as SFGs. Moreover, our sample includes sources classified as RL,AGN, which were identified based on their radio luminosity (i) or $q_{24,\upmu \mathrm m}$ values (v). If an AGN selected using any of the above-discussed criteria is located above $-2\sigma$  dispersion curve of the M\,82 template, the source is categorised as a RQ\,AGN. We assign 418 sources as RL\,AGN and 496 sources as RQ\,AGN out of the total 984 AGN based on our classification method.

The proportion of AGN and SFGs among the 3310 sources classified is presented in Table \ref{tab:class_multi}.
Figure \ref{fig:zhist} presents the redshift distributions of SFGs and AGN recognised using various classification schemes. The open histograms display the distribution of $z_{\rm spec}$ for AGN and SFGs, while the shaded histograms display the distribution of $z_{\rm spec}$ and $z_{\rm ph}$ collectively. The median redshift of SFGs and AGN in our radio catalogue is determined to be 0.89 and 1.46, respectively. We will utilize these classified sources in order to study source properties in the next section.

\subsection{Source Counts for classified sources}\label{sec:sc_classmulti}

We present a detailed analysis of the normalized differential source counts for SFGs, RQ\,AGN and RL\,AGN in the \boot field at 400 MHz. We understand the significance of completeness in determining precise source counts, thus, we have also computed the corrected source counts, taking into account the completeness of all classified sources.

To correct for the completeness of populations in RQ\,AGN, RL\,AGN and SFGs, we followed a similar method as described in section \ref{sec:comp}. The correction factors for these classified sources were derived individually for each category. We again performed a simulation where we injected 1600 SFGs, 500 RQ\,AGN and 500 RL\,AGN (a combination of FRI and FRII) from the SKADS catalogue into our residual RMS map randomly across the field. We performed 100 repetitions of these simulations for each category, i.e., 300 simulations in total.
Thereafter, the correction factor for each of these classifications was derived by employing the method as explained in section \ref{sec:comp} and using eq. \ref{eq:comp}. 

\begin{figure*}
    \centering
    \includegraphics[width=8.25cm]{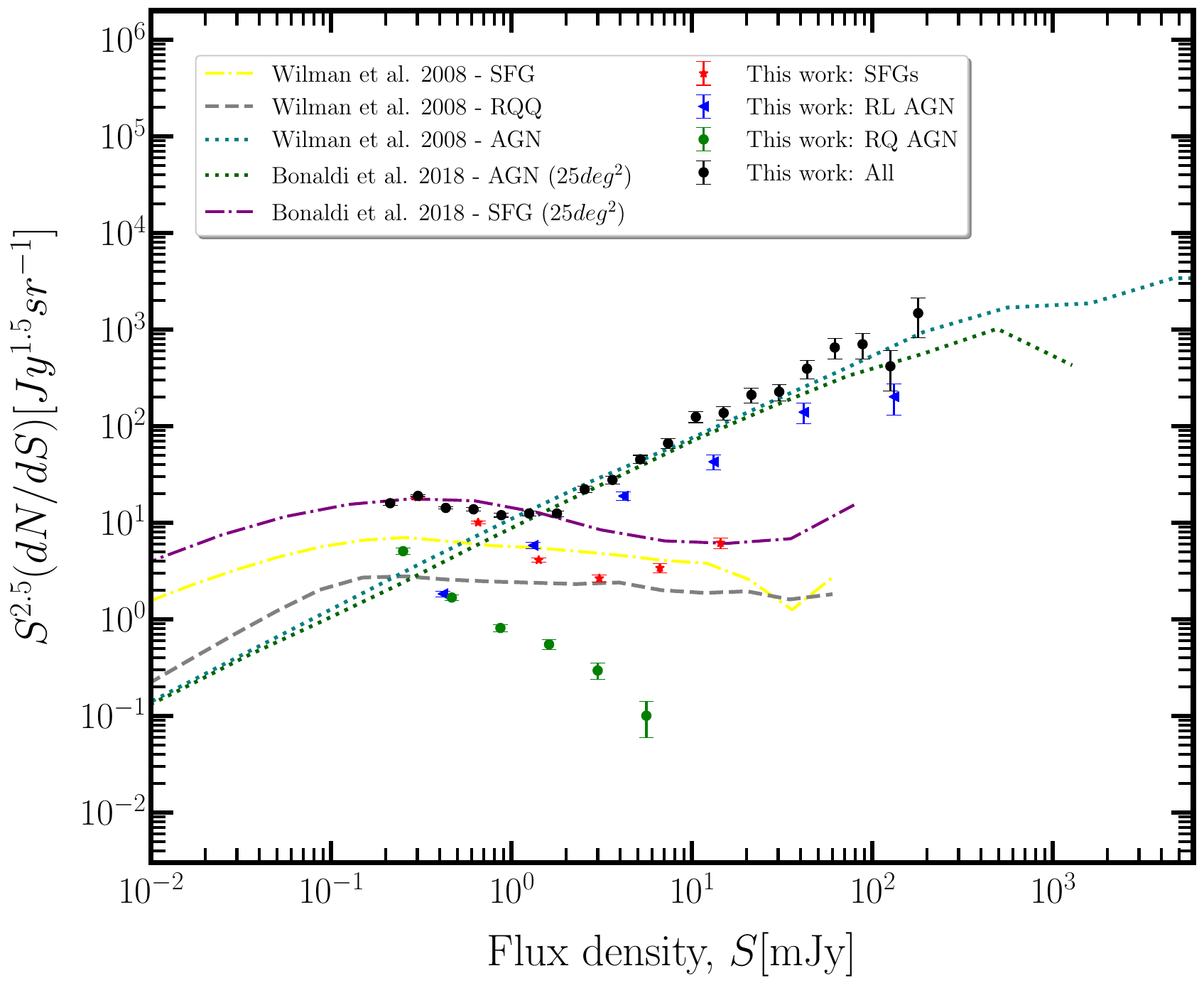}
    \includegraphics[width=8.25cm]{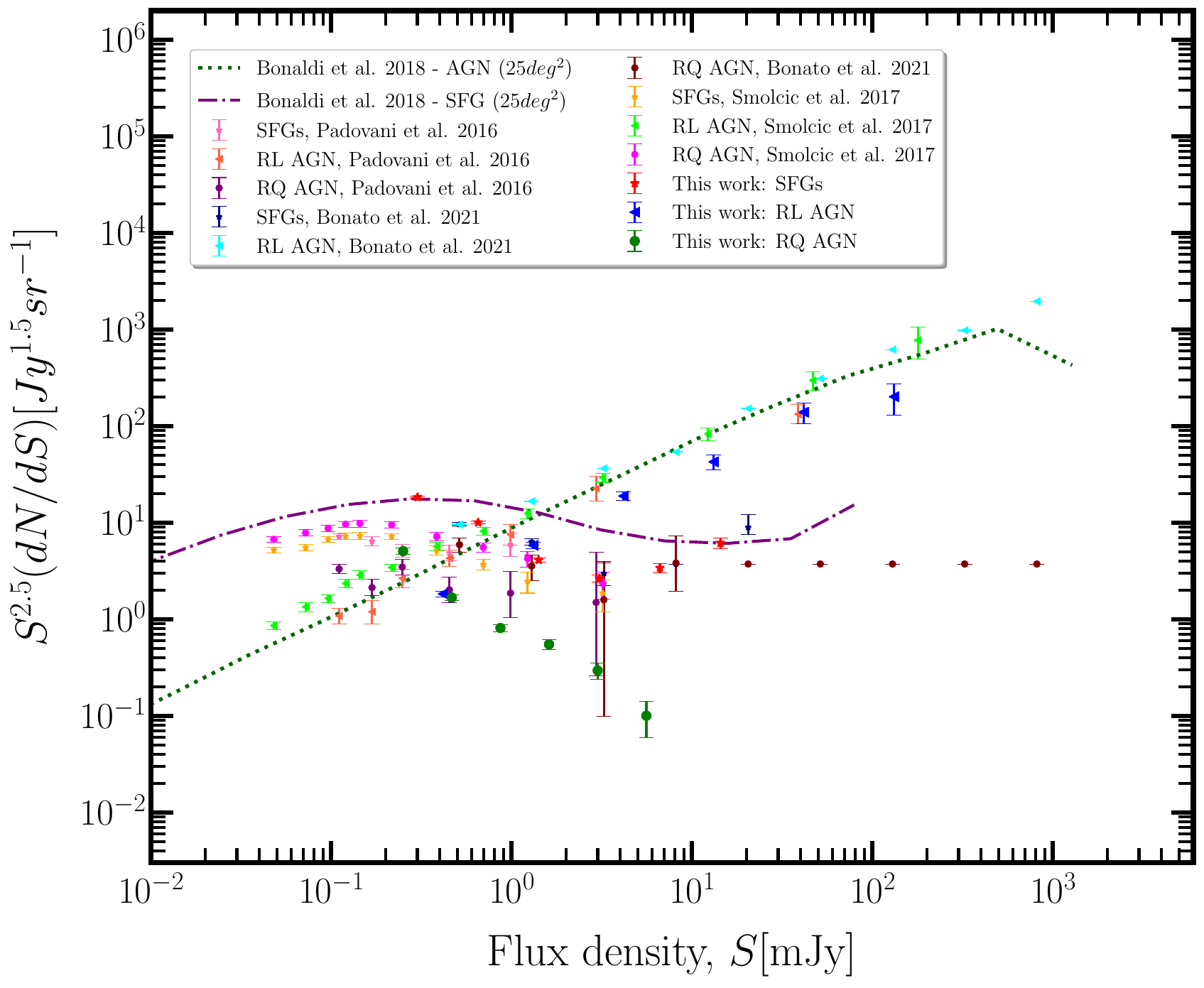}
    \caption{The Euclidean normalised source counts for the SFGs, RQ\,AGN, RL\,AGN and for the combined population shown in red stars, green circles, blue left-triangles and in black circles respectively. In the left-panel, source counts of SFGs, RQQ and AGN modelled by $S^3$-simulations are shown by dashed-dotted yellow, dashed grey and dotted teal lines, while source counts for SFGs and AGN modelled by T-RECS simulations are shown by dash-dotted purple and dotted green lines respectively. In the right-panel, our measures of source counts are shown along with the estimates from \citet{Padovani_2016}, \citet{Smolcic_2017} and \citet{Bonato2021} for the classified sources. }
    \label{fig:sc_class}
\end{figure*}

Our estimates of the normalized differential source count for SFGs (in red), RQ\,AGN (in green) and RL\,AGN (in blue) at 400\,MHz are represented in Fig. \ref{fig:sc_class} and catalogued in Table \ref{tab:sc_class}. In the left-panel of Fig. \ref{fig:sc_class}, we present the comparison of our measures with the models reported by \citet{Wilman_2008} and \citet{Bonaldi_2018}. The model by \citet{Bonaldi_2018}, does not specifically present results on RQ\,AGN, but their SFG source count measures are shown in purple which agrees well with our observations at the lower side of flux bins. 
The study by \citet{Wilman_2008} however underestimates the counts of both SFGs and RQ\,AGN to some extent, in addition to the overall population.

On the right-panel of Fig. \ref{fig:sc_class}, a comparison of our measures of the source counts for the classified population is shown with the existing observations in the literature. The estimates given by \citet{Padovani_2016} and \citet{Bonato2021} at 1.4\,GHz and by \citet{Smolcic_2017} who also scaled their 3\,GHz observations of the COSMOS field to 1.4\,GHz, are used for illustration with scaling of these counts to 400\,MHz using $\alpha =-0.7$. It should be noted that all these studies have used different classification schemes and studied different samples with various approaches. 
We find our measures of the normalised differential source counts to show a similar evolution when compared to the other observations for the classified sources.

From our study at 400\,MHz using uGMRT observations of the \boot field, we also confirm the dominance of SFGs and RQ\,AGN at lower flux densities ($\leq1$\,mJy). Besides, the evolution of source counts for RQ\,AGN and SFGs is somewhat similar, implying that these two populations evolved in a similar manner. However, the counts from the faintest bin for SFGs are slightly higher when compared to the counts for the same population from the literature. It is highly possible that some of the sources are misclassified as SFGs because according to the classification scheme, all the sources with redshift information that were not identified as AGN from any multi-wavelength criteria were identified as SFGs. This could therefore contaminate our sample of SFGs. Further, the counts from the existing observations as mentioned above are centred at 1.4\,GHz and have been scaled using a constant spectral index value. We emphasise the fact that different populations of sources should have different spectral indices that can also vary depending on the redshift and should be considered in the scaling of frequencies. In fact, the counts for the RL\,AGN also show a little lower counts from the existing models and observations, the exact reason for which is not known except for plausible misidentification of the uGMRT sample, however, this population type does dominate at the higher flux density end.

\section{The radio--infrared relations and their evolution}\label{sec:radIR}

Here, we investigate the correlation between radio and infrared emissions by analyzing data from deep observations at 400\,MHz in the \boot field, with a focus on the rest-frame emissions of SFGs along with the AGN population that comprises both the RL\,AGN and RQ\,AGN. The rest-frame radio luminosities ($L_\nu$) were estimated by using eq. \ref{eq:lum}. For estimating the rest-frame IR luminosities, we employ a method that involves modelling the spectral energy distribution (SED) using a single temperature modified-blackbody combined with a truncated mid-IR power law following \citet{Casey_2012} \citep[see][for more details]{Akriti_2022}. We derived the $k$-corrected total IR luminosity ($L_{\rm TIR}$) of a source by integrating its IR SED over the 8--1000$\,\upmu$m rest-frame wavelength. We chose only those sources for which we have IR flux densities at the wavelengths 24, 70, and 160\,$\mu$m from SERVS Data Fusion and at 250, 350 and 500\,$\mu$m from HerMES survey to perform the SED fitting. In this way, we were able to obtain rest-frame TIR luminosities for 390 sources in total from our uGMRT catalogue, of which 327 are SFGs and 63 are AGN. Here, we have not differentiated between RQ\,AGN and RL\,AGN and have considered both populations as simply AGN. Using this fitting and Wien's displacement law, we also measured the dust temperature ($\tdust$), given as, $\tdust = b/\lambda_{\rm peak}$, where, $b = 2.898 \times 10^3 \upmu,\rm{m\,K}$ and $\lambda_{\rm peak}$ is the peak wavelength of the spectrum.

To characterize the radio-IR correlations and their properties, we study the two parameters that quantify these relations: the slope, `$b$', given as $L_{\nu}\propto L_{\rm IR}^b$ and the `$q$' values, defined as, $q=\rm{log}(L_{IR}/L_{\nu})$. For bolometric luminosities, $\qtir$ can be written as \citep{Helou_1985},
\begin{equation}
  q_{\rm TIR} = \log_{10}\left( \frac{L_{\rm TIR}}{3.75 \times 10^{12}~ {\rm [W]}}\right) - \log_{10}\left(\frac{L_\nu}{[\rm{W\,Hz^{-1}}]}\right).
     \label{eq_qIR}
\end{equation}
In this work, we have derived and examined the radio-IR relations using the rest-frame radio luminosity at 400\,MHz ($L_{\rm 400\,MHz}$) with $\ltir$.

\subsection{The radio--TIR relations at 400\,MHz}

\begin{figure}
    \centering
    \includegraphics[width = 8.7cm]{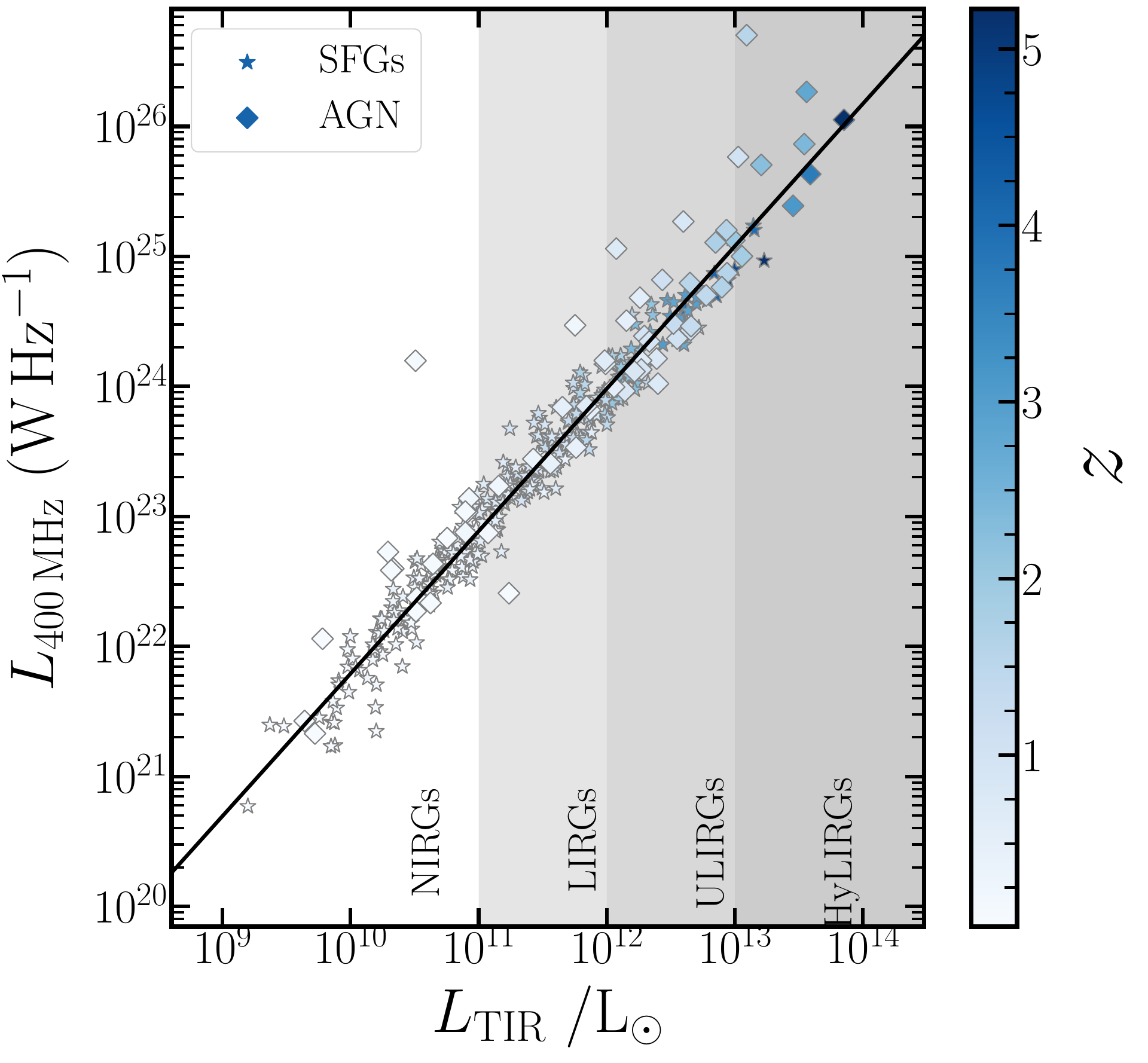}
    \includegraphics[width = 8.6cm]{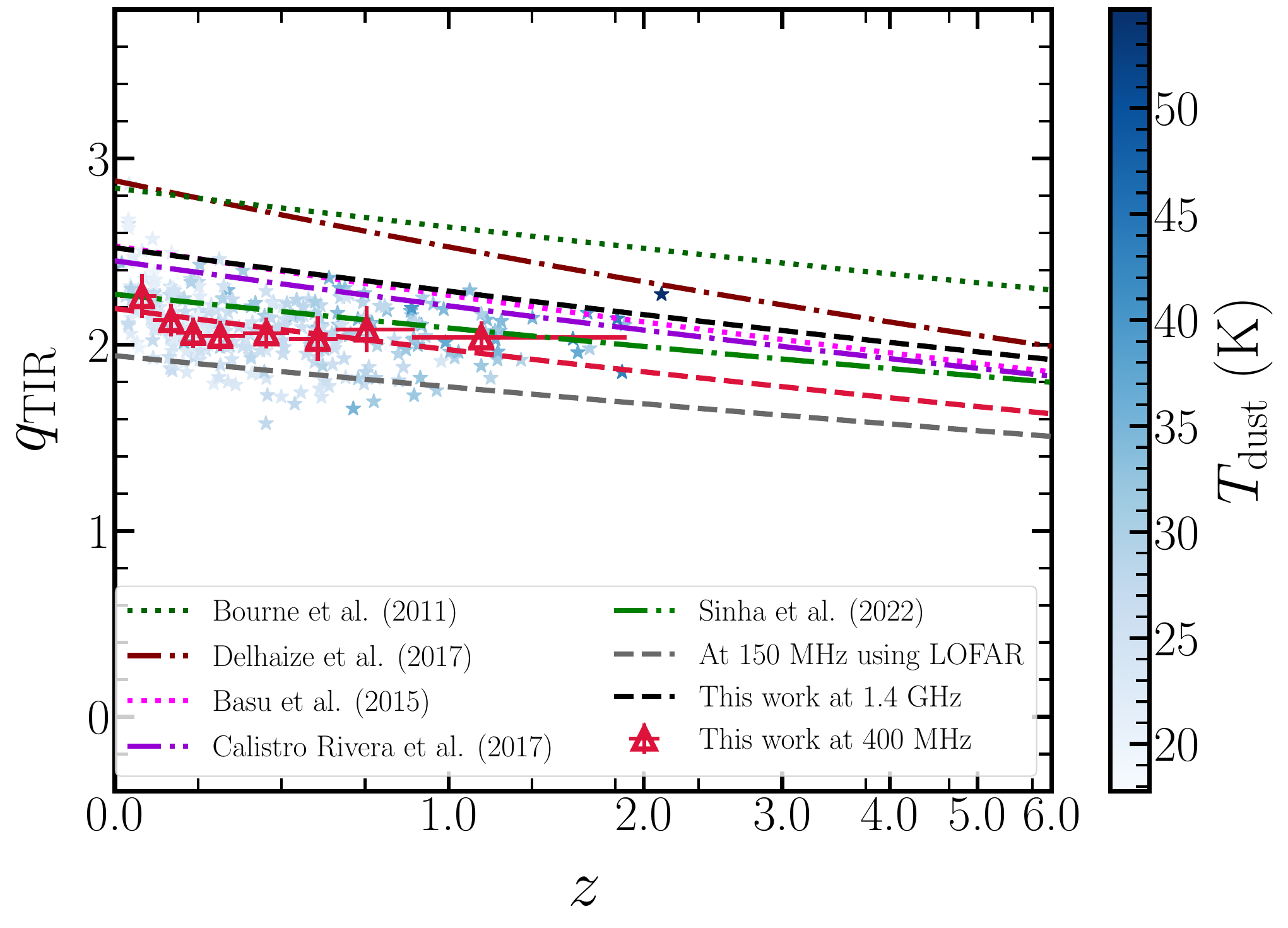}
    \caption{\textit{Top-panel:} The variation of $L_{\rm 400\,MHz}$ with $L_{\rm TIR}$ (between 8--1000$\,\upmu$m). In the plot, the linear regression line indicates the optimal fit for SFGs until a redshift of $z=2$. The data points are represented by star and diamond symbols for SFGs and AGN, respectively. The colour of these symbols indicates their respective redshifts. 
    \textit{Bottom-panel:} Variation of $\qtir$ with $z$. Here, the stars symbol represents SFGs for which $\qtir$ values are measured using $L_{\rm 400\,MHz}$ and $L_{\rm TIR}$. The red open triangles represent the median $\qtir$ values of SFGs up to a redshift of 2, with the power-law fit as a dashed red line.
    The standard deviation of $\qtir$ values is represented by the error bars in the graph for each redshift bin. Additional lines in the graph display the variations in $\qtir$ with $z$ from previous research.}
    \label{fig:rad_ir}
\end{figure}

The top-panel of Fig. \ref{fig:rad_ir} illustrates the variation of $L_{\rm 400\,MHz}$ with the $\ltir$ for SFGs and AGN, represented by star and diamond symbols, respectively. To analyse the correlation between the 400\,MHz luminosity and the infrared luminosity for SFGs, we applied orthogonal distance regression (ODR) to fit the log-log relations of the form $L_{\rm 400,MHz}= aL_{\rm IR}^b$, where a is the normalization factor. To ensure accurate results, we limited the fitting to redshifts less than 2 to avoid any biases from source misidentification or incompleteness in the high-redshift, flux-limited sample. 
We observe a strong $L_{\rm 400\,MHz}$--$\ltir$ correlation for SFGS having Spearman’s rank correlation ($r$) of $0.98$. 
We find these relations to be super-linear with slope $1.10 \pm 0.01$ for SFGs, $1.14 \pm 0.06$ for AGN and $1.12 \pm 0.01$ for the combined population. For further comparison, we also derived the radio-IR relations for these sources using LOFAR fluxes at 150\,MHz. The $L_{\rm 150\,MHz}$--$\ltir$ relations were again found to have super-linear slope values of $1.07 \pm 0.01$, $1.13 \pm 0.07$ and $1.09 \pm 0.01$ for SFGs, AGN and for SFGs$+$AGN combined, respectively. 

The results of our analysis indicate that the radio-IR correlations in our sample are consistent with previous research. These relations have been observed to have a super-linear slope in studies on normal star-forming galaxies by \citet{Bell_2003} and on blue-cloud galaxies by \citet{Basu_2015b} with their measures of slope values as $1.10\pm0.04$ and $1.12\pm0.06$, respectively. Additionally, a sample of 2000 SFGs upto ($z<$0.2) also showed a super-linear slope of $1.11\pm 0.01$ between $L_{\rm 1.4\,GHz}$--$\ltir$, as reported in a study by \citet{molnar2021}. In fact, \citet{Akriti_2022} showed these relations to be again non-linear ($\geq 1.07$) with their analyses on the ELAIS N1 field using radio luminosities at 1.4\,GHz, 400\,MHz  and also for the bolometric luminosities ($L_{\rm RC}$) integrated between 0.1--2\,GHz frequency range. This indicates that the radio-IR correlations for SFGs remain consistent globally up to $z\sim2$ regardless of the method used for classification. Furthermore, the examination and comparison of these relations across various radio frequencies, within the range where synchrotron radiation dominates, does not alter the slope values and the deviation from linearity is consistent across all cases. It thus implies that the observed property of the radio-IR correlations to yield non-linear slope values is inherent in SFGs.

On another note, the shaded region of Fig. \ref{fig:rad_ir} (top-panel), shows the mixed population of galaxies that come into the picture as the redshift increases. These, based on their IR luminosities, are identified as luminous infrared galaxies (LIRGs), ultra-LIRGs (ULIRGs), and hyper-LIRGs (HyLIRGs). Nonetheless, the collective populations satisfy the radio-IR relationships up to high redshift and IR luminosities with negligible deviations from the slope values. The selection bias could also reflect in the results that can be removed with essential deep IR observations ranging from the mid to the far end of the waveband.

\subsection{Variation of $\qtir$ values with $z$}

The `$q$' parameters are frequently employed to investigate the apparent evolution of radio-infrared relations. This can be modeled as a function of $z$, given by, $\qtir \propto \,(1+z)^\upgamma$ \citep[see][for e.g.]{Bourne_2011, Basu_2015b}, where $\upgamma$ is the exponent. 
We applied the same modelling approach to our sample of SFGs up to $z=$2, by dividing them into eight equal-sized redshift bins. Furthermore, the median values of each redshift bin were measured and fitted according to the aforementioned function. Table \ref{tab:qTIR_z_med} lists the median $\qtir$ values and their dispersion corresponding to their median $z$. We have used dispersion for each bin as the errors on $q$ values for robust fitting. We find mild variation in the $\qtir$ values with $z$ using 400\,MHz luminosities as $ q_{\rm TIR} = (2.19 \pm 0.07)\ (1+z)^{-0.15 \pm 0.08}$. The dashed red line represents this in the bottom-panel of Fig. \ref{fig:rad_ir}. The star symbols represent SFGs that are colour-coded based on the $\tdust$. It can be seen that the $\tdust$ variation merely affects the $\qtir$ values, which are obvious as $\qtir$ is measured using the $\ltir$ luminosities that are integrated over the IR spectrum.
The $\qtir$ variation with $z$ for other studies is also shown for comparison and Table \ref{tab:qtir_comp} lists these values.

For further comparison, we derived the $\qtir$ values at 150\,MHz using the LOFAR catalogue \citep{Tasse_2021}. Out of the 327 SFGs, 315 have counterparts in the LOFAR catalogue within a search radius of 3 arcsecs using which we measured the $\qtir$ variation with $z$ as: $ q_{\rm TIR} = (1.94 \pm 0.05)\ (1+z)^{-0.13 \pm 0.06}$. This is shown with a dashed grey line in Fig. \ref{fig:rad_ir} (bottom). Also, now that we have flux densities at 400\,MHz and 150\,MHz for this sample, we calculated the two-point spectral indices using which we extrapolated the 400\,MHz flux densities to measure 1.4\,GHz luminosities. The decrease in $\qtir$ values with $z$ remained, given by, $ q_{\rm TIR} = (2.52 \pm 0.07)\ (1+z)^{-0.14 \pm 0.08}$ and is shown in black dashed lines in the bottom panel of Figure \ref{fig:rad_ir}. In both cases, the evolution of $\qtir$ with $z$ remained consistent with existing observations. Our study's results agree with the findings previously reported in the literature.

\begin{table}
\centering
\caption{The median values of $\qtir$ measured using 400\,MHz luminosities for SFGs in their corresponding redshift bins. The last column marks the standard deviation for the bins. 
The standard deviation for each bin is shown in the last column.}
\begin{tabular}{c c c c}
\hline 
Range of $z$ & Median $z$ & Median ${\qtir}$ & $\sigma$ \\[1ex]
\hline

0.014 - 0.085 & 0.06 & 2.26 $\pm$ 0.01 & 0.12   \\ 
0.086 - 0.136 & 0.12 & 2.13 $\pm$ 0.01 & 0.09   \\ 
0.141 - 0.193 & 0.18 & 2.06 $\pm$ 0.01 & 0.08   \\ 
0.193 - 0.305 & 0.25 & 2.05 $\pm$ 0.01 & 0.08   \\ 
0.310 - 0.430 & 0.37 & 2.06 $\pm$ 0.01 & 0.08   \\ 
0.441 - 0.581 & 0.52 & 2.03 $\pm$ 0.01 & 0.12   \\ 
0.588 - 0.857 & 0.69 & 2.08 $\pm$ 0.01 & 0.12   \\ 
0.861 - 1.883 & 1.14 & 2.04 $\pm$ 0.01 & 0.08   \\ 

\hline
\end{tabular}
\label{tab:qTIR_z_med}
\end{table}

The fact that radio-infrared relations deviate from linearity predicts that the `$q$' parameters should vary as a function of redshift. This is because $q$ in terms of rest-frame radio luminosity and $b$ using the relation,  $L_{\nu}= aL_{\rm IR}^b$, can be written as,
\begin{equation}
\begin{split}
    q & = -\left( \frac{\log_{10}a}{b}\right) + \left(\frac{1-b}{b}\right)\,\log_{10}\,L_\nu\\
\end{split}
\label{eq:q_with_lum}
\end{equation}
using the relation,  $L_{\nu}= aL_{\rm IR}^b$. It is evident that $q$ values depend on the radio luminosity and slope which can be a reason for the non-linearity observed in the radio-IR relation. However, the physical factors responsible for the mild evolution of $q$ parameters with $z$ are yet not clear.

Here, we derived the variation of $\qtir$ with $z$ using $L_{\rm 400\,MHz}$, $L_{\rm 1.4\,GHz}$ and $L_{\rm 150\,MHz}$. We found  a mild but significant evolution of $\qtir$ with $z$ irrespective of the radio frequency, within the range 150\,MHz--1.4\,GHz, This clearly shows that the decrease in $\qtir$ with $z$ is an inherent characteristic of the correlation between radio and IR emissions. Furthermore, the apparent evolution observed in the $q$ values with $z$ at various radio frequencies could also be explained by the evolution of various factors involved in the interstellar medium (ISM), which we will discuss in the next section.

\begin{table*}
    \centering
        \caption{Comparison of the variation of $\qtir$ values with $z$ as studied for various SFGs sample from different fields. The top three rows represent these variations as derived in this work. The remaining rows present the values from the literature and are also shown in Figure \ref{fig:rad_ir} (bottom-panel). The second column represents the frequency at which the $q$ values are measured and their variation with $z$ are presented in the final column. }
    \begin{tabular}{ccc}
         \hline \hline
         SFGs sample from field(s)  &   Frequency (MHz)  & $q_{\rm TIR}$  \\
         \hline
         \boot   &  400 &  $(2.19 \pm 0.07)(1+z)^{-0.15 \pm 0.08}$ \\
        (This work)   &  1400  & $(2.52 \pm 0.07)(1+z)^{-0.14 \pm 0.08}$ \\
                & (scaled using two-point $\alpha$ between 150--400\,MHz) & \\
                &   150  &     $(1.94 \pm 0.05)\ (1+z)^{-0.13 \pm 0.06}$ \\
               &  (LOFAR) & \\
                \hline
         \textit{XMM}-LSS  & 1400   &   $(2.53 \pm 0.02)(1+z)^{-0.16 \pm 0.03}$ \\
         \citep{Basu_2015b} &   (scaled using 325\,MHz) &   \\
         COSMOS, GOODS-N, GOODS-S, ECDFS    &  1400 & $(2.35\pm0.08) (1+z)^{-0.12 \pm 0.04}$\\
         \citep{Magnelli2015}   &   \\
         COSMOS & 1400 & $(2.88\pm0.03)(1+z)^{-0.19\pm0.01}$ \\
         \citep{Delhaize2017}   & (scaled using 3\,GHz) & \\
         \boot  &   1400    &   $(2.45 \pm 0.04)(1+z)^{-0.15 \pm 0.03}$   \\
         \citep{Calistro_2017}  &   (scaled using 150\,MHz) &   \\
         ELAIS\,N1  &   1400    & $(2.86\pm0.04)(1 + z)^{-0.20\pm0.02}$ \\
         \citep{Ocran2020}  &   (scaled using 610\,MHz)    &   \\
         ELAIS\,N1  &   100--2000   &   $(2.27\pm0.03)(1+z)^{-0.12\pm0.03}$  \\
         \citep{Akriti_2022}    & (bolometric)  &   \\
         
         \hline
    \end{tabular}
    \label{tab:qtir_comp}
\end{table*}

\subsection{Importance of ISM in studying the correlations}

The radio and infrared emissions, and hence the $q$ values, are influenced by several physical parameters of the ISM that appear to operate independently. These parameters include, among others, the number density of UV photons and CREs, magnetic field amplification mechanisms, gas and dust densities, energy loss and escape processes of CREs, and star formation history. 
In our previous study \cite{Akriti_2022}, we found that an increase in the magnetic fields within galaxies could be associated with the observed decrease in the $q$ values with $z$. 

From \citet{Basu_2017}, $q$ in terms of the physical characteristics of the interstellar medium for monochromatic radio emission with an assumption of single temperature dust emission can be written as,
\begin{equation}
    q_{\rm IR} = \log_{10}\,\left[\left(\frac{n_{\rm UV}}{n_{\rm CRE}} \right)\,\left(\frac{B_\lambda(T_{\rm dust})\,Q(\lambda,a)}{B_{\rm tot}^{1 - \alpha}\, \nu^\alpha} \right)\right] + C.
    \label{eq:qwithpars}
\end{equation}
where, $n_{\rm UV}$ and $n_{\rm CRE}$ represent the number densities of UV photons and CREs; $B_{\rm \lambda}$ depicts the \textit{Planck} function with $T_{\rm dust}$ as the dust temperature. The absorption coefficient $Q(\lambda, a)$ for dust grains with radius $a$ has the functional form, $Q(\lambda, a) \propto \lambda^\beta$; $B_{\rm tot}$ represents the total magnetic field strength, and $C$ is a normalization factor. The values of these parameters play an important role, as any deviation from them can be reflected in the variation of $q$ values with redshift. Any variations in the $\tdust$ do not significantly affect our results when considering the overall emission of the galaxy and for the bolometric $\qtir$.

In \cite{Akriti_2022}, we demonstrated that variation in $q$ values with $z$ remained even after neglecting the dependence on the radio spectrum by considering the integrated radio emission from galaxies starting from 100\,MHz to 2\,GHz.  In this study, we computed the variation in $q$ values by analyzing the radio flux densities at three different frequencies, namely 400 MHz, 1.4 GHz, and 150 MHz. Our analysis revealed a slight deviation of these values with redshift. Therefore, other factors such as increasing magnetic fields or an increase in the number density of CREs become important in studying the global radio--IR correlations.
\citet{nikla97b} and \citet{Schleicher_Beck_2013} showed that the turbulent dynamo could cause amplified magnetic fields, leading to a relationship between magnetic fields and gas densities, which, in turn, affects the rate of star formation. Thus, for massive galaxies in high redshift, small-scale turbulent dynamo action could explain the increase in magnetic fields with $z$ \citep{Gent_2013,Schober_2016}. In another scenario, the magnetic fields can be seen increasing with $z$ for high luminosity galaxies that are influenced by mergers \citep{Kilerci_2014}. However, there is no strong empirical limitation presently on how magnetic fields evolve as redshift changes. 
Besides, for $q_{\rm IR}$ to decrease, $n_{\rm CRE}$ should increase, and this can be explained as the cosmic SFR density is found to increase upto $z\sim2$ \citep{Madau_2014,Leslie_2020}. Also, $B_{\rm tot}$ increase with $z$ implies a decrease in the energy of CREs, which suggests a rise in the number density of CREs which altogether be responsible for reducing $q$ values at higher $z$.
In recent studies, \cite{delvecchio2021} reported that the reduction in $\qtir$ at higher redshifts may be due to selection bias in flux-limited surveys, which tend to favour more massive galaxies. Another explanation proposed by \citet{molnar2021} suggests that the selection bias that depends on SFR may also contribute to the decline in $\qtir$. 
In our forthcoming paper, we will extend our study to investigate bolometric radio--IR relations in order to understand the cause of the mild variation in $q$ values and the observed non-linear slope values in further detail.

\section{Summary}\label{sec:summary}

In this work, we present our results on uGMRT observations of the \boot field at 300--500\,MHz. The field was observed with seven partially overlapping pointings. These observations span an area of $\sim 5\,\rm{deg}^2$. We used a {\tiny CASA}-based pipeline for the RFI mitigation and direction-independent calibration to make a final mosaic image with $5.0''\times4.9''$ resolution. The achieved RMS noise is $\sim 35\,\mu \rm{Jy\,beam^{-1}}$ resulting in a $5\,\sigma$ sensitivity of $175\,\mu \rm{Jy\,beam^{-1}}$. A total of 3782 sources were extracted using P{\tiny Y}BDSF.

We have compared our catalogue with radio catalogues of the field at different frequencies and found flux accuracy and positional offset to be nearly consistent.
However, we suggest taking into account possible calibration errors by considering an additional 10\,per\,cent uncertainty in the flux values when using this catalogue.
Furthermore, we observed slightly flat spectral indices $\sim -0.63$ when compared to other high and low-frequency catalogues. However, for the WSRT catalogue, we measured a median value of $\alpha = -0.69$. A detailed study of spectral indices and related properties is deferred to future work.

To fully comprehend the physical properties and evolution of various source populations, a multi-wavelength study is necessary. Measuring radio emissions alone is insufficient to uncover their complete nature. The \boot field is a widely-studied extragalactic region that offers a wealth of multi-band ancillary data. This enabled us to classify our sample from uGMRT as RQ\,AGN, RL\,AGN and SFGs. Our identification of AGN from the uGMRT sample, which includes 3310 redshift measurements, is based on a combination of various criteria such as radio luminosity, IRAC colours, spectroscopy, and the ratio of flux densities at 24\,$\mu$m and 1.4 GHz ($q$ values) in addition to classifications obtained from the LoTSS catalogue. In this way, we classified 70\,per\,cent sources as SFGs and 30\,per\,cent sources as AGN.

We present the corrected Euclidean-normalized differential source counts at 400\,MHz and found it to be consistent when compared to existing models and observations. The counts show flattening at the lower flux density regime ($\leq 1\,$mJy) indicating an increase in the fainter source populations. To gain a more thorough understanding, we also derive the source counts for each of the classified groups and found the populations of SFGs and RQ\,AGN to dominate at the fainter flux densities.

Finally, we characterised the properties of the radio--IR correlation for our SFGs and AGN sample by estimating their rest-frame radio and IR luminosities. Here, we have considered only the TIR luminosities integrated in the range 8--1000$\,\upmu$m and studied their correlation with $L_{\rm 400\,MHz}$. We observed a strong super-linear correlation between $L_{\rm 400\,MHz}$ and $\ltir$ with slope, $b= 1.10\pm0.01$ and $r=0.98$. Our analysis reveals that the observed non-linearity in the radio--IR relations is intrinsic to SFGs. Furthermore, we derived the variation of the $q$ parameters with $z$, $ q_{\rm TIR} = (2.19 \pm 0.07)\ (1+z)^{-0.15 \pm 0.08}$. This weak but significant change could be a result of the slight non-linear radio-IR correlations, implying mutual dependence on physical parameters such as magnetic field strengths and their evolution.

\section*{Acknowledgements}

The authors would like to thank the anonymous referee for their comments.
The authors would like to extend their thanks to Aritra Basu and Arnab Chakraborty for their comments and suggestions in the manuscript. AS would further like to thank Aishrila Mazumder for the helpful discussions. AS would like to acknowledge DST for INSPIRE fellowship.
The authors thank the staff of the GMRT that made these observations possible. GMRT is run by the National Centre for Radio Astrophysics of the Tata Institute of Fundamental Research. 
This work is based in part on observations made with the Spitzer Space Telescope, which was operated by the Jet Propulsion Laboratory, California Institute of Technology under a contract with NASA.
This research also made use of Astropy,\footnote{http://www.astropy.org} a community-developed core Python package for Astronomy \citep{astropy:2013, astropy:2018}, NumPy \citep{numpy11}, and Matplotlib \citep{matplotlib07}.
This research has made use of data from HerMES project \href{http://hermes.sussex.ac.uk/}{(http://hermes.sussex.ac.uk/)}. HerMES is a Herschel Key Programme utilising Guaranteed Time from the SPIRE instrument team, ESAC scientists and a mission scientist. The HerMES data was accessed through the Herschel Database in Marseille (HeDaM - \href{http://hedam.lam.fr}{http://hedam.lam.fr)} operated by CeSAM and hosted by the Laboratoire d'Astrophysique de Marseille.

\section*{Data Availability}

The raw data for this study is available in the GMRT archive (\href{https:
//naps.ncra.tifr.res.in/goa/data/search}{https:
//naps.ncra.tifr.res.in/goa/data/search}. The catalogue generated in this study is available with the electronic version of the paper.



\bibliographystyle{mnras}
\bibliography{references} 







\bsp	
\label{lastpage}
\end{document}